\documentclass[12pt]{article} 
\usepackage[margin=1in]{geometry}
\usepackage[longnamesfirst]{natbib}
\usepackage{array,epsfig,fancyheadings,rotating}
\usepackage[unicode]{hyperref}
\usepackage{sectsty, secdot}
\usepackage{diagbox}
\usepackage{amsmath,bm}
\usepackage{amssymb}
\usepackage{amsfonts}
\usepackage{multirow}
\usepackage{amsthm}
\usepackage{color, bm}
\usepackage{graphicx}
\usepackage{subfigure, epstopdf}
\usepackage{setspace}
\usepackage{caption}
\usepackage[ruled]{algorithm2e}
\usepackage{booktabs}
\usepackage{multirow}
\usepackage{threeparttable}
\usepackage{rotating}
\usepackage{fancyhdr}
\usepackage{float}
\usepackage{makecell}

\sectionfont{\fontsize{12}{12pt plus.8pt minus .6pt}\selectfont}
\renewcommand{\theequation}{\thesection\arabic{equation}}
\subsectionfont{\fontsize{12}{12pt plus.8pt minus .6pt}\selectfont}

\newtheorem{thm}{Theorem}

\newtheorem{rem}{Remark}

\newtheorem{pro}{Proposition}

\def\bx{\mathbf{x}}
\def\bz{\mathbf{z}}

\def\bA{\mathbf{A}}

\def\bI{\mathbf{I}}

\def\bS{\mathbf{S}}

\def\0{\mathbf{0}}
\def\1{\mathbf{1}}

\newcommand{\bmu}{\boldsymbol{\mu}}

\newcommand{\bSig}{{\boldsymbol{\Sigma}}}
\newcommand{\bGamma}{{\boldsymbol{\Gamma}}}

\newcommand{\tr}{{\rm tr}}
\newcommand{\E}{{\rm E}}

\newcommand{\var}{{\rm var}}

\newcommand{\T}{{\rm T}}
\newcommand{\rlog}{{\rm log}}

\numberwithin{equation}{section}  

\begin{document}

\captionsetup{font={small}}


\renewcommand{\baselinestretch}{2}

\markright{ \hbox{\footnotesize\rm Statistica Sinica
}\hfill\\[-13pt]
\hbox{\footnotesize\rm
}\hfill }

\markboth{\hfill{\footnotesize\rm FIRSTNAME1 LASTNAME1 AND FIRSTNAME2 LASTNAME2} \hfill}
{\hfill {\footnotesize\rm } \hfill}

\renewcommand{\thefootnote}{}
$\ $\par


\fontsize{12}{14pt plus.8pt minus .6pt}\selectfont \vspace{0.8pc}
\centerline{\large\bf ADAPTIVE TESTS FOR BANDEDNESS OF}
\vspace{2pt} \centerline{\large\bf HIGH-DIMENSIONAL COVARIANCE MATRICES}

\vspace{.4cm} \centerline{Xiaoyi Wang\textsuperscript{1}, Gongjun Xu\textsuperscript{2} and Shurong Zheng\textsuperscript{3}} \vspace{.4cm}
\centerline{\it \textsuperscript{1}Beijing Normal University,  \textsuperscript{2}University of Michigan} \centerline{and \textsuperscript{3}\it Northeast Normal University}
\vspace{.55cm} \fontsize{9}{11.5pt plus.8pt minus.6pt}\selectfont


\begin{quotation}
\noindent {\it Abstract:}
Estimation of the high-dimensional banded covariance matrix is widely used in multivariate statistical analysis. To ensure the validity of estimation, we aim to test the hypothesis that the covariance matrix is banded with a certain bandwidth under the high-dimensional framework. Though several testing methods have been proposed in the literature, the existing tests are only powerful for some alternatives with certain sparsity levels, whereas they may not be powerful for alternatives with other sparsity structures. 
The goal of this paper is to propose a new test for the bandedness of high-dimensional covariance matrix, which is powerful for alternatives with various sparsity levels. The proposed new test also be used for testing the banded structure of covariance matrices of error vectors in high-dimensional factor models. 
Based on these statistics, a consistent bandwidth estimator is also introduced for a banded high dimensional covariance matrix.
Extensive simulation studies and an application to a prostate cancer dataset from protein mass spectroscopy are conducted for evaluating the effectiveness of the proposed adaptive tests blue and bandwidth estimator for the banded covariance matrix.

\vspace{9pt}
\par
\end{quotation}\par

\def\thefigure{\arabic{figure}}
\def\thetable{\arabic{table}}

\renewcommand{\theequation}{\thesection.\arabic{equation}}

\fontsize{12}{14pt plus.8pt minus .6pt}\selectfont

\setcounter{section}{0} 
\setcounter{equation}{0} 

\lhead[\footnotesize\thepage\fancyplain{}\leftmark]{}\rhead[]{\fancyplain{}\rightmark\footnotesize\thepage}

\section{Introduction\label{Section-1}}
Statistical testing of covariance matrix plays an important role in multivariate and high-dimensional statistical analysis, for example, in principle component analysis, multivariate regression analysis, and factor analysis (see \citealt{anderson2003introduction,bai2008limit,johnstone2001distribution,cai2011adaptive,fan2015power}).

Many researchers have studied testing high-dimensional covariance matrices from different aspects.
One aspect is to test $H_{01}: \bSig=\bSig_0$, where $\bSig$ is the population covariance matrix, and $\bSig_0$ is a given positive definite matrix.
For instance, \cite{ledoit2002some} proposed a robust statistic for testing $H_{01}$ based on the Frobenius norm under the Gaussian assumption.
Without the Gaussian assumption, \cite{bai2009corrections} developed a corrected likelihood ratio test (LRT) for the identity test issue when the dimension $p$ of data is smaller than the sample size $n$. 
\cite{jiang2012likelihood} studied the asymptotic distribution of the corrected LRT for normal random vectors when $p/n\rightarrow y\in (0,1]$. Later,
\cite{wang2013identity} redefined the above statistics and introduced two tests that can accommodate data with unknown mean and non-Gaussian distribution.
Moreover, \cite{chen2010tests} and \cite{cai2013optimal} constructed sum-of-squared type tests through U-statistics as $n, p\rightarrow \infty$.
Another aspect is to test $H_{02}: \bSig=c\bSig_0$, where $c$ is an unknown positive number.
For the sphericity testing, \cite{ledoit2002some} and \cite{chen2010tests} proposed sum-of-squared type statistics, the former directly used the sample covariance matrix $\bS_n$ in substitution for $\bSig$ and the latter adopted unbiased U-statistics. Furthermore, 
\cite{wang2013sphericity} developed a corrected LRT $(p<n)$ and John's test. 
\cite{jiang2013central}  extended the corrected LRT to the case of $p/n\rightarrow y\in(0,1]$, and \cite{li2016testing} proposed a quasi-LRT allowing $p/n\rightarrow \infty$.
In addition, researchers also are interested in testing a general linear structure of covariance matrices. \cite{zheng2019hypothesis}
studied the problem of testing $H_{03}: \bSig=\theta_1\bA_1+\cdots+\theta_K\bA_K$,
where $\theta_1,...,\theta_K$ are unknown parameters and $\bA_1,...,\bA_K$ are known basis matrices. 
Furthermore, \cite{zhong2017tests} introduced an adjusted goodness-of-fit test that examines a broad range of covariance structures to assess the adequacy of specified covariance structures.

 In this paper, we are interested in testing the banded structure of covariance matrices which has numerous applications in biological science, climate, econometrics, finance, etc (see, \citealt{andrews1991heteroskedasticity,ligeralde1995band}).
For instance, in high-dimensional data analysis, a popularly used covariance matrix estimation method is banding or tapering the sample covariance matrix \citep[e.g.,][]{bickel2008regularized}. Although the large-sample consistency of the corresponding estimators has been established for the covariance matrices in the ``bandable" class, it remains questionable whether or not the underlying covariance matrix belongs to the ``bandable" class. 
 The considered hypothesis testing on the banded structure of covariance matrices will provide a practical statistical guideline to this issue.

Several methods for testing the bandedness of the high-dimensional covariance matrix have been proposed.
In particular, \cite{qiu2012test} developed a test using a linear combination of sample U-statistics by collecting the sum-of-squares of all covariance differences between the null and alternative hypotheses.
The above sum-of-squares-type test is powerful against dense alternatives because there are many nonzero components in the covariance differences between the null and alternative hypotheses. However, it is not powerful when the alternative is sparse.
To address this problem, \cite{cai2011limiting} proposed a maximum-type statistic by capturing the maximum componentwise sample covariance difference for multivariate normal random vectors.
\cite{shao2014necessary} restudied the above statistic and suggested using chi-square distribution instead of the type I extreme distribution to improve the convergence rate of maximum-type statistic. Furthermore, \cite{xiao2013asymptotic} relaxed the normality assumption based on the normalized maximum componentwise sample covariance difference.
The maximum-type test is powerful when the alternative is sparse, while less powerful for dense alternatives.
In practice, however, it is often unclear whether the alternative hypothesis is dense, sparse, and in-betweens.
What is more, neither type of these tests are powerful when the alternative hypothesis is denser or less sparse, which will be shown in Section \ref{Section-3}.

Motivated by this, we propose two adaptive tests based on a series of unbiased U-statistics for the banded structure of high-dimensional covariance matrices, following the idea of the adaptive test in \cite{xu2016adaptive} and \cite{he2021asymptotically}. Our contributions are as follows.

\begin{itemize}
\item [(i)]
We derive the joint asymptotic distribution of the series of U-statistics under the null hypothesis. Furthermore, we show that the U-statistics are asymptotically independent and jointly normally distributed under certain regularity conditions.
\item [(ii)] 
We establish the asymptotic distribution of the series of finite order U-statistics under a local alternative hypothesis. Furthermore, we compare the power performance of these U-statistics and show the consistency of these tests.
\item [(iii)]
We proposed two adaptive tests by combining the $p$-values of the U-statistics, and their consistency will be guaranteed by those single U-statistics.
These adaptive tests will select the test with the most significant result and yield high powers under a broad spectrum of alternative hypothesis scenarios.
\item [(iv)] 
We provide an adaptive estimator for the bandwidth of the high-dimensional banded covariance matrix and establish its consistency.
\end{itemize}
The rest of the paper is organized as follows. In Section \ref{Section-2}, we first introduce a series of U-statistics, deriving their joint asymptotic distributions under the null and local alternative hypotheses. 
Furthermore, we propose two adaptive tests and reduce the computation burden of the U-statistics. At last, we also present a bandwidth estimator for the banded covariance matrix and show its consistency. In Section \ref{Section-3}, extensive simulation studies are demonstrated. In Section \ref{Section-4}, we analyze a prostate dataset to demonstrate our procedures. Section \ref{Section-5} concludes the paper and discusses the potential work of this paper.
Main technical proofs and more simulation results are relegated to the Supplementary Material.

\section{New Test Methods\label{Section-2}}
Let $\bx_i=(x_{i, 1},\dots,x_{i, p})^{\T}$, for $i=1,\dots,n$, be independent and identically distributed (i.i.d.) samples from a $p$-dimensional population $\bx=(x_1,\dots,x_p)^{\T}$ with mean vector $\bmu=(\mu_1,\dots,\mu_p)^{\T}$ and covariance matrix $\bSig=(\sigma_{j_1j_2})_{p\times p}$.
The population covariance matrix $\bSig=(\sigma_{j_1j_2})_{p\times p}$ is said to be banded if there exists an integer $k\in\{0,\dots,p-2\}$ such that
$\sigma_{j_1j_2}=0$ for $|j_1-j_2|>k$. The smallest $k$ such that $\bSig$ is banded is called the bandwidth of $\bSig$.
Let $\boldsymbol{B}_k(\bSig)=(\sigma_{j_1j_2}\1_{\{|j_1-j_2|\leq k\}})_{p\times p}$ be the banded version of $\bSig$ with bandwidth $k$, where $\1_{\{\cdot\}}$ is an indicator function. When $k=0$, $\boldsymbol{B}_0(\bSig)$ is the diagonal version of $\bSig$.
In this paper, we are interested in testing
\begin{align}
\label{H0}
H_{k,0}: \bSig=\boldsymbol{B}_k(\bSig)  \quad  v.s. \quad  H_{k,1}:  \bSig\neq \boldsymbol{B}_k(\bSig)
\end{align}
for a certain positive integer $k$. We further rewrite the hypothesis testing (\ref{H0}) as
$$
H_{k,0}: \mathcal{E}= \0 \quad  v.s. \quad  H_{k,1}: \mathcal{E} \neq \0,
$$
where $\mathcal{E}=\{\sigma_{j_1j_2}:  k<|j_1-j_2|< p\}$  is the parametric set we are interested in.

\subsection{A series of U-statistics\label{Section-2-1}}
Motivated by \cite{he2021asymptotically},
we consider a series of measurements of $\mathcal{E}$, defined by 
$||\mathcal{E}||_a =\left[\sum\limits_{k<|j_1-j_2|< p}(\sigma_{j_1j_2})^a\right]^{1/a},$
and intend to construct test statistics that are powerful against $||\mathcal{E}||_a$ for different finite positive integer $a$.
Since $\E(x_{i_1,j_1}x_{i_1,j_2}-x_{i_1,j_1}x_{i_2,j_2})=\sigma_{j_1j_2}$ for $1\leq i_1\not=i_2 \leq n$, we propose the U-statistic
\begin{eqnarray*}
\mathcal{U}(a)\!=\!
\sum\limits_{k<|j_1-j_2|< p}(P^n_{2a})^{-1}\sum\limits_{1\leq i_1\neq \cdots \neq i_{2a}\leq n}\prod_{l=1}^a
(x_{i_{2l-1},{j_1}}x_{i_{2l-1},{j_2}}-x_{i_{2l-1},{j_1}}x_{i_{2l},{j_2}})
\end{eqnarray*}
as an unbiased estimator of $||\mathcal{E}||_a^a$, where $P^n_{2a}= {n!}/{(n-2a)!}$ denotes the number of $2a$-permutations of $n$.
A straightforward calculation shows that 
\begin{eqnarray}
\label{U1}
\mathcal{U}(a)&=&\sum\limits_{k<|j_1-j_2|< p}\sum\limits_{c=0}^a\binom{a}{c}(-1)^c(P_{a+c}^n)^{-1}
\sum\limits_{1\leq i_1\neq \cdots \neq i_{a+c}\leq n}\nonumber\\
&&\prod_{l=1}^{a-c}(x_{i_{l},{j_1}}x_{i_{l},{j_2}})\prod_{s=a-c+1}^{a}x_{i_{s},{j_1}}\prod_{t=a+1}^{a+c}x_{i_{t},{j_2}}.
\end{eqnarray}
 The form of $\mathcal{U}(a)$ in (\ref{U1}) plays an essential role in deriving the theoretical properties of our proposed statistics.
Specifically, to obtain the expression in (\ref{U1}), we define $\varphi_{j_1 j_2}=\E(x_{i,j_1}x_{i,j_2})$ and
$\sigma_{j_1 j_2}=\E[(x_{i,j_1}-\mu_{j_1})(x_{i,j_2}-\mu_{j_2})]=\varphi_{j_1 j_2}-\mu_{j_1}\mu_{j_2}$. For any finite positive integer $a$, we have
\begin{eqnarray}
\sum\limits_{k<|j_1-j_2|< p}\sigma_{j_1 j_2}^a&=&\sum\limits_{k<|j_1-j_2|< p}(\varphi_{j_1 j_2}-\mu_{j_1}\mu_{j_2})^a \nonumber\\
&=&\sum\limits_{k<|j_1-j_2|< p}\sum\limits_{c=0}^a\binom{a}{c}(-1)^c\varphi_{j_1 j_2}^{a-c} \mu_{j_1}^c\mu_{j_2}^c.
\end{eqnarray}
Since $x_{i,j}$ and  $x_{i,j_1}x_{i,j_2}$ are the unbiased estimators of $\mu_{j}$ and $\varphi_{j_1 j_2}$, respectively, for $1\leq i_1\neq\cdots\neq i_{a+c}\leq n$, it follows that
$\E(\prod_{l=1}^{a-c}x_{i_l,j_1}x_{i_l,j_2}\prod_{s=a-c+1}^{a}$\\$x_{i_s,j_1}\prod_{t=a+1}^{a+c}x_{i_t,j_2})
=\varphi_{j_1 j_2}^{a-c}\mu_{j_1}^c\mu_{j_2}^c.$
Thus, we obtain the expression (\ref{U1}).

\begin{rem}
\label{rem2}
If we only consider the term of $c=0$ in $(\ref{U1})$, we have
\begin{eqnarray}
\label{tilde-U}
\tilde{\mathcal{U}}(a)=(P_{a}^n)^{-1}\sum\limits_{k<|j_1-j_2|< p}
\sum\limits_{1\leq i_1\neq \cdots \neq i_{a}\leq n}\prod_{l=1}^{a}(x_{i_{l},{j_1}}x_{i_{l},{j_2}}),
\end{eqnarray}
which will be shown to be a leading term of $(\ref{U1})$ under certain regularity conditions specified  in Section \ref{Section-2-2} and be used for our theoretical analysis in the Supplementary Material.
\end{rem}

\subsection{Asymptotic properties of U-statistics under null hypothesis\label{Section-2-2}}
Before deriving the theoretical properties of U-statistics under the null hypothesis, we first introduce some notations as follows: $u_{n,p}=o(v_{n,p})$, if $\limsup_{n,p\rightarrow\infty}|u_{n,p}/v_{n,p}|=0$; $u_{n,p}=\Theta(v_{n,p})$, if $0<\liminf_{n,p\rightarrow\infty}|u_{n,p}/v_{n,p}|\leq\limsup_{n,p\rightarrow\infty}|u_{n,p}/v_{n,p}|<\infty$ and 
\begin{eqnarray}
\label{pi}
\mathop{\Pi}\nolimits_{j_1,...,j_t}=\E[(x_{1,j_1}-\mu_{j_1})\cdots(x_{1,j_t}-\mu_{j_t})].
\end{eqnarray}
We assume the following regularity conditions in our analysis:
\begin{description}
\item [{\it Condition 1. }]
$\mathop{\lim}\limits_{p\rightarrow \infty}\mathop{\max}\limits_{1\leq j\leq p}\E[(x_j-\mu_j)^8]<\infty$ and
$\mathop{\lim}\limits_{p\rightarrow \infty}\mathop{\min}\limits_{1\leq j\leq p}\E[(x_j-\mu_j)^2]>0$.

\item [{\it Condition 2. }]
A sequence of random variables $\bz=\{z_j, j \geq 1\}$ is said to be $\alpha$-mixing if $\lim\limits_{s\rightarrow \infty}\alpha_{\bz}(s)=0$, where $\alpha_{\bz}(s)=\sup_{t\geq 1}\{|P(A\cap B)-P(A)P(B)|: A\in \mathcal{F}_1^t, B\in \mathcal{F}_{t+s}^\infty\}$
with $\mathcal{F}_a^b$ being the $\sigma$-algebra generated by $\{z_a, z_{a+1},...,z_b\}$.
Under $H_0$, we assume $\bx$ is $\alpha-$mixing with $\alpha_{\bx}(s)\leq M\delta^s$, where $\delta\in (0,1)$ and $M$ is some positive constant.

\end{description}

The regularity conditions are similar to those of Theorem 2.1 in \cite{he2021asymptotically}, which studied the problem of testing the diagonality of the covariance matrix, a special case of \eqref{H0} with $k=0$. 
Specifically, Condition 1 requires that the eighth marginal moments of $\bx$ are uniformly bounded from above and the second marginal moments are uniformly bounded from below. 
Condition 2 prescribes weak dependence among the column components of random vector $\bx=(x_1,\dots,x_p)^{\T}$ in $\alpha$-mixing type, which is satisfied when $\bx$ is $m$-dependent random vector or Gaussian distributed random vector with banded covariance matrix. There are several strong mixing conditions, such as $\phi$-mixing, $\psi$-mixing, $\rho$-mixing, and $\beta$-mixing. In the classical theory, these five strong mixing conditions have emerged as the most prominent ones, and the $\alpha$-mixing condition is the weakest one among those (e.g. \citet{bradley2005basic}). 
The $\alpha$-mixing condition has also been imposed in many research works, such as \cite{xu2016adaptive,chen2019two}. 
In our work, the $\alpha$-mixing condition to these mixture moments $\E(\prod\nolimits_{t=1}^s x_{j_t})$ for $2\leq s \leq 8$ ensures the asymptotic independence of different finite order U-statistics. 
In addtion, \cite{bai1996effect} assumed the independent component structure $\bx=\bmu+\bGamma \bz$ to describe the weak dependence among the components of $\bx$. The random vector $\bx$ is $\alpha$-mixing when $\bGamma$ is a $p\times p$ upper triangular matrix with $\gamma_{j_1,j_2}=0$ for $j_1-j_2>k$.

\begin{thm}
\label{the1}
Under Conditions 1, 2 
and $H_{k,0}$, for any finite positive integers $a_1,...,a_m$, we have
\begin{eqnarray}
\bigg(\frac{\mathcal{U}(a_1)}{\sigma(a_1)},\cdots,\frac{\mathcal{U}(a_m)}{\sigma(a_m)}\bigg)^{\mathop{\T}}\xrightarrow{D}\mathcal{N}(0,I_m),~~n, p\rightarrow\infty,
\end{eqnarray}
where
\begin{eqnarray}
\label{var-real}
\sigma^2(a) = \var[\mathcal{U}(a)] = (P_a^n)^{-1}a!
\sum\limits_{\mbox{\tiny$\begin{subarray}{c}
k<|j_1-j_2|< p\\
k<|j_3-j_4|< p\end{subarray}$}}(\mathop{\Pi}\nolimits_{j_1,j_2,j_3,j_4})^a+o(n^{-a}p^2)
\end{eqnarray}
with $\prod\nolimits_{j_1,j_2,j_3,j_4}$ defined in (\ref{pi}).
\end{thm}
Because $\sigma^2(a)$ is unknown,   we obtain the following theorem where $\sigma^2(a)$ is replaced by an estimator $\hat{\sigma}^2(a)$ provided in (\ref{var-est}).
To ensure the consistency of $\hat{\sigma}^2(a)$, the following Condition 3 will be needed.

\begin{description}
\item [{\it Condition 3. }]
For a finite positive integer $a$, $\mathop{\lim}\limits_{p\rightarrow \infty}\mathop{\max}\limits_{1\leq j\leq p}\E[(x_j-\mu_j)^{8a}]<\infty$. 
\end{description}

\begin{thm}
\label{the2}
Under Conditions 1, 2  and $H_{k,0}$, for any positive integers $a_1,...,a_m$ satisfying Condition 3, we have
\begin{eqnarray}
\bigg(\frac{\mathcal{U}(a_1)}{\hat{\sigma}(a_1)},\cdots,\frac{\mathcal{U}(a_m)}{\hat{\sigma}(a_m)}\bigg)^{\mathop{\T}}\xrightarrow{D}\mathcal{N}(0,I_m),~~n, p\rightarrow\infty,
\end{eqnarray}
where $\hat{\sigma}^2(a)/\sigma^2(a)\xrightarrow{P}1$ and
\begin{align}
\label{var-est}
\hat{\sigma}^2(a) =  2(P_a^n)^{-2}a! \!\!\!\!\!\!\!\!
\sum\limits_{\mbox{\tiny$\begin{subarray}{c}
k<|j_1-j_2|< p\\
k<|j_3-j_4|< p\\
|j_1-j_3|\leq k,
|j_2-j_4|\leq k
\end{subarray}$}} \!\!\!\!\!\!
\sum\limits_{1\leq i_1\neq \cdots \neq i_{a}\leq n}
\!\!\! \prod\nolimits_{l=1}^a [(x_{i_l,j_1}-\bar{x}_{j_1})\cdots(x_{i_l,j_4}-\bar{x}_{j_4})].
\end{align}
\end{thm}
Theorem \ref{the2} shows that $ {\mathcal{U}(a_1)}/{\hat{\sigma}(a_1)}$, $\dots$, $ {\mathcal{U}(a_m)}/{\hat{\sigma}(a_m)}$ are {\sl\it asymptotically independent and normally distributed}.
The theoretical results in Theorems \ref{the1}-\ref{the2} extend those in 
   \cite{he2021asymptotically} from testing the diagonality of the covariance matrix to testing a general banded structure, which is often of practical interest in high-dimensional covariance matrix estimation. Technically, the general banded structure makes the analysis more involved, the details are presented in the supplementary file.

\begin{rem} 
\label{Remark-2}
For an extreme case, as an even number $a\rightarrow\infty$, we have 
$$||\mathcal{E}||_a =\left[\sum\limits_{k<|j_1-j_2|< p}\sigma_{j_1j_2}^a\right]^{1/a}\longrightarrow||\mathcal{E}||_{\infty}=\max\limits_{k<|j_1-j_2|< p}|\sigma_{j_1j_2}|.$$ 
Thus, the performance of the statistic $~\mathcal{U}(a)$ would be similar to the maximum-type statistics when the even order $a$ is large. This phenomenon was also observed in \cite{xu2016adaptive} and \cite{he2021asymptotically}. 
\cite{he2021asymptotically} provided the asymptotic independence between the finite order U-statistics and infinite order U-statistic (maximum-type statistic) when the components of the random vector $\bx$ are uncorrelated. We also expect a similar result under certain regular conditions in our setting. However, it is challenging to establish the asymptotic joint distribution of the maximum-type statistic and finite order U-statistics because the banded covariance structure is much more complicated than the i.i.d. case due to the dependence, which was pointed out in \cite{cai2011limiting}. We will investigate it in the future. 
\end{rem}

\subsection{Power analysis}
\label{Section-2-2-1}
In this section, we investigate the limiting distributions of the series of U-statistics under the local alternative hypothesis $H_{k,A}: \bSig=\bSig_A$, which is described in Condition 4. For a given bandwidth $k$, we denote the set of locations of the signals by $J_{A} = \{(j_1,j_2): \sigma_{j_1j_2}\neq0, k<|j_1-j_2|<p, j_1,j_2=1,\dots,p\}$, and the cardinality of $J_{A}$ by $|J_{A}|$ which stands for the sparsity level of $\bSig_A$. The sparsity level of the alternative hypothesis decreases as $|J_{A}|$ increases. 
We introduce two conditions for presenting the asymptotic distribution under the local alternative hypothesis.
\begin{description}
\item [{\it Condition 4. }]
Assume $|J_{A}|=o(p^2)$ and for any $(j_1, j_2)\in J_{A}$, $|\sigma_{j_1j_2}|=\Theta(\rho)$, where $\rho=\sum\nolimits_{(j_1,j_2)\in J_{A}}|\sigma_{j_1j_2}|/|J_{A}|$.
\item [{\it Condition 5. }]
For $t\leq 8$, we assume that there exists constant $\kappa$ such that $\mathop{\Pi}\nolimits_{j_1,\cdots,j_t}=\kappa \E(\prod\nolimits_{k=1}^t z_{j_k})$, where $1\leq j_1,\dots,j_t\leq p$ and $(z_1,\dots,z_p)^{\rm{T}}\sim\mathcal{N}(\mathbf{0},\bSig_A)$.
\end{description}

\begin{thm}
\label{the3}
Under Conditions 1, 4 and 5, for any positive integers $a_1,\dots,a_m$, if $\rho = O(|J_{A}|^{-1/a_t}p^{1/a_t}n^{-1/2})$ for $t=1,\dots,m$, we have
\begin{eqnarray*}
\bigg(\frac{\mathcal{U}(a_1)-\E_A[\mathcal{U}(a_1)]}{\sigma_A(a_1)},\cdots,\frac{\mathcal{U}(a_m)-\E_A[\mathcal{U}(a_m)]}{\sigma_A(a_m)}\bigg)^{\mathop{\T}}\xrightarrow{D}\mathcal{N}(\mathbf{0},\bI_m),~~n, p\rightarrow\infty,
\end{eqnarray*}
where $\E_A[\mathcal{U}(a)]=\sum\limits_{(j_1,j_2)\in J_{A}} \sigma_{j_1j_2}^a$ and
$\sigma_A^2(a) \simeq  2(P_a^n)^{-1}a!\kappa^a 
\sum\limits_{\mbox{\tiny$\begin{subarray}{c}
|j_1-j_3|\leq k\\
|j_2-j_4|\leq k
\end{subarray}$}}\sigma_{j_1j_3}^a\sigma_{j_2j_4}^a.
$
with the order $\Theta(n^{-a}p^2)$.
\end{thm}

Given the asymptotic properties under the local alternatives, the power function of a single U-statistic $\mathcal{U}(a)$ is
\begin{eqnarray}
\label{power-func}
\beta(a) = P\bigg(\frac{\mathcal{U}(a)}{\sigma(a)}>z_{1-\alpha}\bigg|H_{k,A}\bigg) \rightarrow \Phi\bigg(-z_{1-\alpha}+\frac{\E_A[\mathcal{U}(a)]}{\sigma_A(a)}\bigg),
\end{eqnarray}
where $z_{1-\alpha}$ and $\Phi(\cdot)$ are the $(1-\alpha)$th quantile and cumulative distribution function of standard normal distribution, respectively. The signal-to-noise ratio $\mbox{SNR}_{a}=\E_A[\mathcal{U}(a)]/\sigma_A(a)$ plays an important role in affecting the power performance of the U-statistic $\mathcal{U}(a)$. 
For any finite order $a\in\mathcal{I}$, we define the corresponding average standardized signal as $\bar{\rho}_a=\sum\nolimits_{(j_1,j_2)\in J_{A}} n^{a/2}\sigma_{j_1j_2}^a/|J_{A}|$. Based on Theorem \ref{the3}, the asymptotic power function $\beta(a)\rightarrow 1$ if $p^{-1}|J_A|\bar{\rho}_a\rightarrow\infty$ because $\E_A[\mathcal{U}(a)]=\sum\nolimits_{(j_1,j_2)\in J_{A}} \sigma_{j_1j_2}^a$ and $\sigma_A(a)$ are of order $n^{-a/2}p$. In other words, if $\bar{\rho}_a$ is of order higher than $p|J_{A}|^{-1}$, $\beta(a)\rightarrow 1$ as $n\rightarrow\infty$.

Another attractive work is to investigate the relationship between the order of the U-statistic with the highest asymptotic power and the sparsity level $|J_{A}|$. We give a criterion to compare the power performance of two finite order U-statistics $\mathcal{U}(a_1)$ and $\mathcal{U}(a_2)$. We call $\mathcal{U}(a_1)$ is better than $\mathcal{U}(a_2)$ if $\rho_{a_1}<\rho_{a_2}$ when they attain the same asymptotic power. 
Particularly, we consider a special case where the signal strength is fixed at the same level, 
$\sigma_{j_1j_2}=\rho>0$ for $(j_1,j_2)\in J_{A}$ and $\sigma_{j_1j_3}=\sigma_{j_2j_4}=\nu>0$ for 
$|j_1-j_3|\leq k$ and $|j_2-j_4|\leq k$. 
In this case, 
$$\mbox{SNR}_{a}\simeq\frac{\sum\nolimits_{(j_1,j_2)\in J_{A}} \sigma_{j_1j_2}^a}{\{2(P_a^n)^{-1}a!\kappa^a
\sum\nolimits_{\mbox{\tiny$\begin{subarray}{c}
|j_1-j_3|\leq k\\
 |j_2-j_4|\leq k
\end{subarray}$}}(\sigma_{j_1j_3}\sigma_{j_2j_4})^a\}^{1/2}}.$$
Hence, the power function
$\beta(a)\rightarrow \Phi\Big(-z_{1-\alpha}+\frac{|J_{A}|\rho^a}{\sqrt{2a!\kappa^{a}}\nu^a n^{-a/2}p'}\Big)$, where $p'=(2k+1)\sqrt{(p-k-1)(p-k)}$.
\begin{eqnarray}
\label{rho-a}
\rho_a = (Mp'/|J_{A}|)^{1/a}(a!)^{1/2a}\kappa^{1/2}\nu n^{-1/2},
 \end{eqnarray}
achieves the asymptotic power $\Phi(-z_{1-\alpha}+M/\sqrt{2})$ of $\mathcal{U}(a)$, where $M$ is some constant.
Proposition \ref{power-compare} establishes the relationship between the sparsity level and the order of the U-statistic.
\begin{pro}
\label{power-compare}
For a given bandwidth $k$, under the special case described above, given $n, p, |J_{A}|$ and $M$, by considering (\ref{rho-a}) as a function of integer order $a$, we have
\begin{itemize}
\item [(i).] when $|J_{A}|\geq Mp'$, the minimum of $\rho_a$ is achieved at $a=1$;
\item [(ii).] when $|J_{A}|< Mp'$, the minimum of $\rho_a$ is achieved at some $a$, which increases as $Mp'/|J_{A}|$ increases.
\end{itemize}
\end{pro}
When $|J_{A}|\geq Mp'$, the alternative is very dense, $\mathcal{U}(1)$ is the most powerful test. When $|J_{A}|< Mp'$, as $Mp'/|J_{A}|$ increases, the sparsity level of the alternative hypothesis increases, the U-statistic with larger order will perform better. This result is consistent with the analysis in \cite{he2021asymptotically}, and we extend their result to the banded covariance matrix setting.

\subsection{Two adaptive testing procedures}
\label{Section-2-3}

For the proposed family of U-statistics, $\mathcal{U}(a)$ is powerful against the alternative with large  $||\mathcal{E}||_a^a = \sum\limits_{k<|j_1-j_2|< p}\sigma_{j_1j_2}^a$. The power performance of $\mathcal{U}(a)$ is determined by the sparsity and the strength of signals. The test with a smaller order $a$ would be preferred for a   denser alternative. For example, $\mathcal{U}(1)$ is the most powerful one when the alternative is very dense as shown in Section \ref{Section-3-1}.
In practice, it is often unclear which test statistic should be chosen because the true alternative is usually unknown.
Therefore,  motivated by the idea in \cite{xu2016adaptive} and \cite{he2021asymptotically}, we develop the adaptive tests by combining the information from U-statistics with different orders, which would yield high powers against various alternatives.

We propose two adaptive tests:  one based on the minimum combination method and the other based on Fisher's method.
Suppose that we have a candidate set $\mathcal{I}=\{a_1,\dots,a_m\},~|\mathcal{I}|=m$, where $|\mathcal{I}|$ denotes the cardinality of $\mathcal{I}$. Let $p_a$ be the $p$-value of test $\mathcal{U}(a)$ as
$p_a=2 (1-\Phi(|\mathcal{U}(a)/\hat{\sigma}(a)|)).$

\noindent {\bf Minimum combination method}:
Rejecting $H_0$ if $p_{\text{adpUmin}}<\alpha$ where
\begin{eqnarray}
\label{Proce1}
p_{\text{adpUmin}}=1-(1-T_{\text{adpUmin}})^{|\mathcal{I}|},~~T_{\text{adpUmin}} = \min\limits_{a\in\mathcal{I}}p_a,
\end{eqnarray}
with the nominal significance level $\alpha$.
The type I error of the minimum combination method can be controlled by
$P(p_{\text{adpUmin}}<\alpha) = P(T_{\text{adpUmin}}<p_{\alpha}^*) \rightarrow \alpha,$
where $p_{\alpha}^*=1-(1-\alpha)^{1/|\mathcal{I}|}$ and the asymptotic independence of
${\mathcal{U}(a_1)}/{\hat{\sigma}(a_1)}$, $\dots$, $ {\mathcal{U}(a_m)}/{\hat{\sigma}(a_m)}$ are used.

\noindent{\bf Fisher's method}:
We have $T_{\text{adpUf}} = -2\sum_{a\in \mathcal{I}}\rlog~ p_a \xrightarrow{D} \chi^2_{2|\mathcal{I}|},$ where
$\chi^2_{2|\mathcal{I}|}$ is distributed as a chi-square distribution with degrees of freedom $2|\mathcal{I}|$.
We reject $H_0$ if $p_{\text{adpUf}}<\alpha$ with
\begin{eqnarray}\label{Proce2}
p_{\text{adpUf}}=1-\Psi(T_{\text{adpUf}}),
\end{eqnarray}
where $\Psi(\cdot)$ is the cumulative distribution function of $\chi^2_{2|\mathcal{I}|}$ with degrees of freedom $2|\mathcal{I}|$.

\begin{rem} 
For the two adaptive statistics, we have:
\begin{itemize}
\item[(i)] $P(T_{\text{adpUmin}}=\min\limits_{a\in\mathcal{I}}p_a<p_{\alpha}^*)\geq P(p_a<p_{\alpha}^*)\rightarrow \Phi\big(-z_{1-p_{\alpha}^*}+\frac{\E_A[\mathcal{U}(a)]}{\sigma_A(a)}\big)$
\item[(ii)] $P(T_{\text{adpUf}}=-2\sum\limits_{a\in \mathcal{I}}\rlog p_a>c_{1-\alpha})\geq P(-2\rlog p_a>c_{1-\alpha})\rightarrow \Phi\big(-z_{1-c^*_\alpha}+\frac{\E_A[\mathcal{U}(a)]}{\sigma_A(a)}\big)$, where $c^*_\alpha=e^{-\frac{1}{2}c_{1-\alpha}}$ and $c_{1-\alpha}$ is the $(1-\alpha)$th quantile of $\chi^2_{2|\mathcal{I}|}$.
\end{itemize}
The asymptotic power of the proposed adaptive tests converge to 1 as $n\rightarrow \infty$ if there is a U-statistic $\mathcal{U}(a)$ satisfies average standardized signal $\bar{\rho}_a$ is of order higher than $p|J_A|^{-1}$ with $\bar{\rho}_a=\sum\nolimits_{(j_1,j_2)\in J_A} n^{a/2}\sigma_{j_1j_2}^a/|J_A|$.
\end{rem}

\begin{rem}
Our proposed adaptive tests are versatile in the sense that they can adapt to the unknown sign and sparsity level of the signal set $\mathcal{E}$ under the alternative hypotheses. Their performances depend on the selection of the order set $\mathcal{I}$. The U-statistics with odd order may lose their power performance quickly because the differences across the sign of the elements in $\mathcal{E}$ leads to the cancellation of positive and negative $\sigma_{j_1j_2}^a$. In this case, we suggest using U-statistics with even order to construct the adaptive tests. However, the U-statistics with odd order are still more suitable when the elements in $\mathcal{E}$ are all in the same direction. For example, $\mathcal{U}(1)$ is a representative of the burden tests based on genotype pooling or collapsing which has been discussed in \cite{morgenthaler2007strategy, li2008methods, pan2014powerful}. Without the information of the directions of signals, we suggest using both odd and even order U-statistics. Furthermore, the theoretical arguments on power analysis and extensive simulation studies indicate that the order of the best U-statistic will increase as the sparsity level decrease. To address sparse alternative hypotheses, we select the biggest order to be $6$ because the performance of $\mathcal{U}(6)$ is good enough compared to the maximum-type statistic, as shown in the first figure in Figure \ref{fig1}. The discussions in \cite{xu2016adaptive} and \cite{he2021asymptotically} also support our suggestion.
\end{rem}

\begin{rem}
It is exciting to study whether the proposed U-statistics can achieve the optimal detection/testing boundary at different sparsity levels. Such a problem for the U-statistics, however, differs from existing studies (e.g. \citet{donoho2004higher}) due to the differences among the studied testing problems, and needs new theoretical development to handle the dependence structure of the banded covariance matrix. When testing $\bSig=\bI$, \cite{cai2013optimal} showed that $\mathcal{U}(2)$ is rate optimal in terms of the Frobenius norm for the testable region and the non-testable region. It would be interesting to extend this result to the U-statistics with different orders for testing the banded covariance matrix, and we would like to leave that as a future study.
\end{rem}


\subsection{Simplifying computation}
\label{Section-2-5}
The costs of directly calculating $\mathcal{U}(a)$ in (\ref{U1}) and $\hat{\sigma}^2(a)$ in (\ref{var-est}) are as expensive as $O(n^{2a}p^{2})$ and $O(n^{a}p^{2})$, respectively. To reduce the computational cost, we mainly employ the Algorithm 1 proposed in \cite{he2021asymptotically} by changing the input $s_{i,l}$, that reduces the computation cost across $i$ from $O(n^{2a})$ or $O(n^{a})$ to $O(n)$.

When $\E(x_{ij})$ is known, we assume $\E(x_{ij})=0$ without loss of generality. In this case, $\mathcal{U}(a)$ degenerates into $\tilde{\mathcal{U}}(a)$ in (\ref{tilde-U}). 
{\bf (1).} In computing $\tilde{\mathcal{U}}(a)$, we specify $s_{i,l} = x_{i,{j_1}}x_{i,{j_2}}$ in Algorithm 1, where $i=1,\dots,n$ and $l\in\mathcal{L} = \{(j_1,j_2): k < |j_1- j_2| < p\}$.
{\bf (2).} Similarly, we compute $\hat{\sigma}^2(a)$ with $s_{i,l} = (x_{i,j_1}-\bar{x}_{j_1})(x_{i,j_2}-\bar{x}_{j_2})(x_{i,j_3}-\bar{x}_{j_3})(x_{i,j_4}-\bar{x}_{j_4})$, where $i=1,\dots,n$ and $l\in\mathcal{L} = \{(j_1,j_2,j_3,j_4): k< |j_1- j_2| < p, k < |j_3- j_4| < p, |j_1-j_3|\leq k, |j_2-j_4|\leq k\}$. 

When $\E(x_{ij})$ is unknown, we present the following proposition to discuss the computation of $\mathcal{U}(a)$.
\begin{pro}
\label{computation}
The forms of $\mathcal{U}(a)$ with different order $a$ are as follows.
\begin{itemize}
\item [(i)] When $a=1$,
$\mathcal{U}(1)=\!\!\!\!\!
\sum\limits_{k<|j_1-j_2|< p}\!\!\!\!\!
\big\{n^{-1}\sum\limits_{i=1}^n x_{i,{j_1}}x_{i,{j_2}}\!-(P_{2}^n)^{-1}
\big(\sum\limits_{i_1=1}^n x_{i_1,{j_1}}\!\!\sum\limits_{i_2=1}^n x_{i_2,{j_2}}$\\$
-\sum\limits_{i=1}^n x_{i,{j_1}}x_{i,{j_2}}\big)\big\}.$
\item  [(ii)] When $a=2$,
$\mathcal{U}(2)=\!\!\!\!\!\sum\limits_{k<|j_1-j_2|< p}\!\!\!\!\big\{(P_{2}^n)^{-1}U_0(2)-2(P_{3}^n)^{-1}U_1(2)+(P_{4}^n)^{-1}$\\$U_2(2)\big\},$
with
$U_0(2) = (\sum\limits_{i=1}^n x_{i,{j_1}}x_{i,{j_2}})^2-\sum\limits_{i=1}^n (x_{i,{j_1}}x_{i,{j_2}})^2$,
$U_1(2) = (\sum\limits_{i=1}^n x_{i,{j_1}}$\\$x_{i,{j_2}})U_{11}(2)-U_{12}(2)-U_{13}(2)$ and 
$U_2(2) = \prod\limits_{s=1}^2 \{(\sum\limits_{i=1}^n x_{i,{j_s}})^2-(\sum\limits_{i=1}^n x^2_{i,{j_s}})\}-2U_0(2)-4U_1(2)$ with
$U_{11}(2) = (\sum\limits_{i=1}^n x_{i,{j_1}})(\sum\limits_{i=1}^n x_{i,{j_2}})-\sum\limits_{i=1}^n x_{i,{j_1}}x_{i,{j_2}}$,
$U_{12}(2) = (\sum\limits_{i=1}^n x^2_{i_,{j_1}}x_{i,{j_2}})(\sum\limits_{i=1}^n x_{i,{j_2}})-\sum\limits_{i=1}^nx^2_{i_,{j_1}}x^2_{i_,{j_2}}$,
$U_{13}(2) = (\sum\limits_{i=1}^n x_{i_,{j_1}}x^2_{i,{j_2}})(\sum\limits_{i=1}^n x_{i,{j_2}})-\sum\limits_{i=1}^nx^2_{i_,{j_1}}x^2_{i_,{j_2}}$.

\item  [(iii)] When $a\geq 3$, let
$\mathcal{U}_c(a)=(P_{a}^n)^{-1}\sum\limits_{k<|j_1-j_2|< p}
\sum\limits_{1\leq i_1\neq \cdots \neq i_{a}\leq n}\prod_{l=1}^{a}(x_{i_{l},{j_1}}-\bar{x}_{j_1})(x_{i_{l},{j_2}}-\bar{x}_{j_2}).$
Under the Conditions 1, 2, 3 and $H_0$, if $a$ is odd, $p=o(n^{1+a/2})$; if $a$ is even, $p=o(n^{a/2})$. Then
$\{\mathcal{U}(a)-\mathcal{U}_c(a)\}/\sigma(a)\xrightarrow{P}0$.
\end{itemize}
\end{pro}
When $a=1, 2$, we directly compute $\mathcal{U}(a)$ using Proposition \ref{computation}.(i)--(ii).
When $a\geq3$, $\mathcal{U}(a)$ can be replaced by $\mathcal{U}_c(a)$ induced by Proposition \ref{computation}.(iii).
We compute $\mathcal{U}_c(a)$ with Algorithm 1 by setting $s_{i,l} = (x_{i,{j_1}}-\bar{x}_{j_1})(x_{i,{j_2}}-\bar{x}_{j_2})$, where $i=1,\dots,n$ and $l\in\mathcal{L} = \{(j_1,j_2): k < |j_1- j_2| < p\}$.

\subsection{Adaptive bandwidth estimation}
\label{S2-6}
Based on the by-product of the studied U-statistics, we propose a method to estimate the bandwidth parameter of the high dimensional banded covariance matrix $\bSig$. Our method is motivated by \cite{qiu2012test}.
To facilitate the illustration, we define some notations. For a given bandwidth parameter $k$, we denote the corresponding statistic $\mathcal{U}(a)$ in $(\ref{U1})$ as $\mathcal{U}_{a,k}$, its asymptotic standard deviation $\sigma(a)$ and asymptotic standard deviation estimator $\hat{\sigma}(a)$ as $\sigma_{a,k}$ and $\hat{\sigma}_{a,k}$, respectively. Following \cite{qiu2012test}, we consider a banded covariance matrix with true bandwidth $k_0$. We define $\mathcal{T}_{a,k}=n^{-1}\mathcal{U}_{a,k}/\hat{\sigma}_{a,k}$, and rewrite it as $\mathcal{T}_{a,k}=\mathcal{T}_{a,k,1}+\mathcal{T}_{a,k,2}$, where 
$$\mathcal{T}_{a,k,1}=n^{-1}\frac{\mathcal{U}_{a,k}-\mu_{a,k}}{\sigma_{a,k}}\frac{\sigma_{a,k}}{\hat{\sigma}_{a,k}}
\quad\mbox{and}\quad
\mathcal{T}_{a,k,2}=n^{-1}\frac{\mu_{a,k}}{\sigma_{a,k_0}}\frac{\sigma_{a,k_0}}{\sigma_{a,k}}\frac{\sigma_{a,k}}{\hat{\sigma}_{a,k}}.$$
Since $\{\mathcal{U}_{a,k}-\mu_{a,k}\}/\sigma_{a,k}$ is stochastically bounded and $\sigma_{a,k}/\hat{\sigma}_{a,k}\xrightarrow{P}1$, then $\mathcal{T}_{a,k,1}=O_p(n^{-1})$. In addtion, since both $\sigma_{a,k_0}$ and $\sigma_{a,k}$ are order of $\Theta(n^{-a/2}p)$, $\mathcal{T}_{a,k,2}$ is determined by 
$
n^{-1}\frac{\mu_{a,k}}{\sigma_{a,k_0}}=\frac{n^{-1}\sum\nolimits_{k<|j_1-j_2|<p}\sigma_{j_1j_2}^a}{[(P_a^n)^{-1}a!
\sum\nolimits_{\mbox{\tiny$\begin{subarray}{c}
k_0<|j_1-j_2|< p\\
k_0<|j_3-j_4|< p\end{subarray}$}}(\mathop{\Pi}\nolimits_{j_1,j_2,j_3,j_4})^a]^{1/2}}.
$
Specially, if all the signs of the covariances $\sigma_{j_1,j_2}$ are positive with $|j_1-j_2|\leq k$, 
it can be checked that $n^{-1}\mu_{a,k}/\sigma_{a,k_0}>0$ for $k<k_0$ and $n^{-1}\mu_{a,k}/\sigma_{a,k_0}=0$ for $k\geq k_0$.
It inspires us to consider an estimator based on the difference between successive statistics $d_{a,k} = \mathcal{T}_{a,k}-\mathcal{T}_{a,k+1}$ for a given finite order $a\in\mathcal{I}$. We multiply $n^{\delta}$ on $\mathcal{T}_{a,k}$ with a small positive $\delta\in(0,1)$ to increase the magnitude of $\mathcal{T}_{a,k,2}$ and ensure that $\mathcal{T}_{a,k,1}$ converges to 0 in probability with a quick rate. For any $a\in\mathcal{I}$, we define $d_{a,k}^{\delta}=n^{\delta}(\mathcal{T}_{a,k}-\mathcal{T}_{a,k+1})$, and come up with the bandwidth estimator 
\begin{align}
\hat{k}_{a,\delta,\theta} = \min\{k: |d_{a,k}^{\delta}|<\theta\}.
\end{align}
By combining the influence induced by different orders, we finally propose an adaptive bandwidth estimator 
$
\hat{k}_{\delta,\theta}= \max\limits_{a\in \mathcal{I}} \hat{k}_{a,\delta,\theta}.
$
We conduct a simulation study to illustrate the motivation of $\hat{k}_{\delta,\theta}$ in the appendix of supplmentary file.
We also present the consistency of the bandwidth estimator $\hat{k}_{\delta,\theta}$.
\begin{pro}
\label{the4}
Under Conditions 1, 2, 3 and $\mathop{\lim\inf_{n}}\{\inf_{k<k_0}(\mu_{a,k}-\mu_{a,k+1})\}$\\$>0$, for any banded covariance matrix with bandwidth $k_0$, then $\hat{k}_{\delta,\theta}-k_0\xrightarrow{P}0$, for any $\theta>0$ and $\delta\in(0,1)$.
\end{pro}
In Proposition \ref{the4}, $\mathop{\lim\inf_{n}}\{\inf_{k<k_0}(\mu_{a,k}-\mu_{a,k+1})\}>0$ excludes the case that there exists a zero sub-diagonal followed by nonzero sub-diagonals as one moves away from the main diagonal. The performance of the adaptive estimator $\hat{k}_{\delta,\theta}$ may be affected by the tuning parameters $\theta$ and $\delta$. As pointed out in \cite{qiu2012test}, the multiplier $n^{\delta}$ leads to $\theta$ being ``free ranged" as long as $\theta>0$. We suggest practitioners to choose $\delta=0.5$ to trade off the converge rate of $\mathcal{T}_{a,k,1}$ and the performance of $\mathcal{T}_{a,k,2}$. The performance of our adaptive bandwidth estimator $\hat{k}_{\delta,\theta}$ with Monte Carlo simulation studies is presented in Section \ref{Sim-bandwidth} .
 
\section{Simulation Study\label{Section-3}}

In this section, we conduct comprehensive simulation studies to evaluate the performance of our adaptive tests and estimator.
We generate $n$ random vectors
$\bx_i = (x_{i1},\dots,x_{ip})^{\T}$ from two populations:
(i). multivariate normal distribution: $N(\0,\bSig)$;
(ii). multivariate t distribution with seven degrees of freedom: $t_7(\0,\bSig)$, where $\bSig=\bGamma\bGamma^{\T}$. We choose the index set $\mathcal{I}=\{1,\dots,6\}$.

\subsection{Adaptive Testing Methods\label{Section-3-1}}
For $a\in\mathcal{I}$, let ``$\mathcal{U}(a)$" denote the testing procedure with the rejection region $\{\bx_1,\dots,\bx_n: |\mathcal{U}(a)|/\hat{\sigma}(a)>q_{1-\alpha/2}\}$
and $q_{1-\alpha/2}$ being the $(1-\alpha/2)100\%$ quantile of $N(0, 1)$. Denote ``adpUmin" and ``adpUf" as our proposed testing procedures in
(\ref{Proce1}) and (\ref{Proce2}). We also compare ``adpUmin", ``adpUf" with
``$\mathcal{U}(1)$", ``$\mathcal{U}(2)$", ``$\mathcal{U}(3)$", ``$\mathcal{U}(4)$", ``$\mathcal{U}(5)$", ``$\mathcal{U}(6)$", ``QC" in \cite{qiu2012test} and ``XW" in \cite{xiao2013asymptotic}.
We take $n=100$, $p=50,100,200,400,600,800,1000$ to present the empirical sizes, and 
$n = 100$, $p = 600, 1000$ to investigate the empirical powers.
The population covariance matrix $\bSig=\bGamma\bGamma^{\T}$ varies under three different settings as follows.
{Before introducing the settings, we use $J_{|J_A|,k}$ to present a set of $|J_A|$ random positions $(j_1,j_2)$ which satisfy $j_2-j_1>k$.

{\it Setting 1.} Let
$\bGamma = (\gamma_{j_1 j_2})_{p\times p}$, when $ j_2-j_1=1$, $\gamma_{j_1j_2}=1$; when $(j_1,j_2)\in J_{|J_A|,1}$, $\gamma_{j_1j_2}=\rho$, otherwise, $\gamma_{j_1j_2}=0$.
We investigate the empirical sizes with $|J_A|=0$, and the empirical powers by varying the signal magnitude $\rho\in (0,1)$ and the sparsity level $|J_A|=2,400,1200,2400$.
In this setting, the bandwidth $k=1$ under $H_{k,0}$.

{\it Setting 2}. Let
$\bGamma = (\gamma_{j_1 j_2})_{p\times p}$, when $j_2-j_1=1$, $\gamma_{j_1 j_2}=0.8$; when $j_2-j_1=2$, $\gamma_{j_1 j_2}=0.6$; when $(j_1,j_2)\in J_{|J_A|,2}$, $\gamma_{j_1 j_2}= \rho_A$, otherwise $\gamma_{j_1 j_2}= 0$.
We investigate the empirical sizes with $|J_A|=0$, and the empirical powers by varying the signal magnitudes $\rho_A$ which are generated from $\mbox{Unif}(0,2\rho)$ with $\rho\in(0,1)$ and the sparsity level $|J_A|=2,400,1200,2400$.
In this setting, the bandwidth $k=2$ under $H_{k,0}$. 

{\it Setting 3.} Let
$\bGamma = (\gamma_{j_1 j_2})_{p\times p}$, when $j_2-j_1=1,\dots,5$, $\gamma_{j_1 j_2}= 0.6$; when $j_2-j_1=5+a$, $\gamma_{j_1 j_2}= \rho$, otherwise $\gamma_{j_1 j_2}=0$.
We investigate the empirical sizes with $a=0$, and the empirical powers by varying the signal magnitude $\rho$ and the sparsity level $a=1, 3, 6, 10, 15, 25$. 

The simulation replication times are $1000$ and the nominal test level $\alpha=5\%$.
Table \ref{size1} presents the empirical sizes with multivariate normal populations for different combinations of $n$ and $p$ under Setting 1. The simulation results show that
the empirical sizes of all the compared tests are close to the nominal level $5\%$.  
Table \ref{size4} exhibits the empirical sizes with multivariate t random samples with degrees of freedom seven.
The empirical sizes of these single U-statistic tests and our proposed adaptive tests are still close to $5\%$.
However, the empirical sizes of ``QC" and ``XW" are far away from the nominal level.
Figure \ref{fig1} summarizes the empirical powers under Setting 1 for multivariate normal random vectors.
The empirical power profiles in Figure \ref{fig1} show that
\begin{itemize}
\item
For extremely sparse alternatives with $|J_A|=2$, $\mathcal{U}(6)$ performs well;
\item
For moderately sparse alternatives with $|J_A|=400$, $\mathcal{U}(4)$ performs well;
\item
For dense alternatives with $|J_A|=1200$ and $2400$, $\mathcal{U}(1)$ and $\mathcal{U}(2)$ perform well;
\item
When $|J_A|$ increases, the empirical power of ``QC" also increases, but the empirical power of ``XW" decreases.
Nonetheless, our proposed two testing procedures ``adpUmin" and ``adpUf" always maintain high empirical powers regardless of small $|J_A|$ or large $|J_A|$. It also appears that ``adpUf" generally performs better than ``adpUmin".
\end{itemize}

In summary, the adaptive tests either achieved the highest powers or were close to the test with the highest power in any setting, indicating their good performance across a wide range of situations. Due to space limitation, we present other simulation results in Section S9.1 of the Supplementary Material. The conclusions are similar to those of Tables \ref{size1}--\ref{size4} and Figure \ref{fig1}.

\subsection{Adaptive Bandwidth Estimator\label{Sim-bandwidth}}
We compare our proposed adaptive bandwidth estimator (Adaptive) with the estimator (BLa) discussed in \cite{bickel2008regularized} and the fixed estimator (QC) in \cite{qiu2012test}. For bandwidth estimation, we set the parameters $\delta=0.5$ and $\theta=0.06$ in our proposed estimator $\hat{k}_{\delta,\theta}$ and QC estimator.  The parameter setting of BLa estimator is chosen to be the same as theirs. We set $n=100, 200$ and $p=50, 200, 400, 600, 1000$ in the following Models 1--4 with true bandwidth $k=2, 5, 10, 15$, respectively.

{\it Model 1.} Let
$\bGamma = (\gamma_{j_1 j_2})_{p\times p}$, when $j_2-j_1=1$, $\gamma_{j_1 j_2}=0.8$; when $j_2-j_1=2$, $\gamma_{j_1 j_2}=0.6$, otherwise, $\gamma_{j_1 j_2}= 0$.

{\it Model 2.} Let
$\bGamma = (\gamma_{j_1 j_2})_{p\times p}$, when $j_2-j_1=1,\dots,5$, $\gamma_{j_1 j_2}= 0.6$, otherwise, $\gamma_{j_1 j_2}=0$.

{\it Model 3.} Let
$\bGamma = (\gamma_{j_1 j_2})_{p\times p}$, when $j_2-j_1=1,\dots,5$, $\gamma_{j_1 j_2}= 0.2$; $j_2-j_1=6,\dots,10$, $\gamma_{j_1 j_2}= 0.4$, otherwise $\gamma_{j_1 j_2}= 0$. 

{\it Model 4.} Let 
$\bGamma = (\gamma_{j_1 j_2})_{p\times p}$, when $j_2-j_1=1,\dots,10$, $\gamma_{j_1 j_2}= 0.2$; when $j_2-j_1=11,\dots,15$, $\gamma_{j_1 j_2}= 0.4$, otherwise, $\gamma_{j_1 j_2}= 0$. 

Table \ref{band-estimation-normal} reports the average empirical bias and standard deviations with the innovations from normal distribution based on 100 replications. From Table \ref{band-estimation-normal}, we observe that our proposed estimator performs well compared to BLa since the smaller bias and standard deviation. The bias and standard deviation of the BLa estimator increase as the dimension $p$ gets larger owing to the inappropriate estimation of the covariance matrix under the high-dimensional setting. 
Similar results of these estimators with $t_7(\0,\bSig)$ are presented in Tabel 5 in Section S9.2 of the supplementary file. 

\begin{table}[h]
\caption{\rm Empirical sizes under Setting 1 for $N(\0,\bSig)$ and $n = 100$ (in percentage).}
\renewcommand{\arraystretch}{0.9}\doublerulesep 0.01pt\tabcolsep 0.18in
\begin{center}
\label{size1}
\scalebox{1}{
\begin{tabular}{cp{0.5cm}<{\centering}p{0.5cm}<{\centering}p{0.5cm}<{\centering}p{0.5cm}<{\centering}p{0.5cm}<{\centering}p{0.5cm}<{\centering}p{0.5cm}<{\centering}}
\hline
$p$ & 50 & 100 & 200 & 400 & 600 & 800 &1000 \cr
\hline
$\mbox{adpUmin}$ &4.70 &6.40 &6.70 &7.20 &5.60 &5.10 &4.30  \cr
$\mbox{adpUf}$ &5.60 &6.80 &6.30 &6.90 &5.80 &5.70 &4.80  \cr
$\mathcal{U}(1)$ &4.60 &5.70 &4.90 &5.60 &6.10 &5.00 &5.00  \cr
$\mathcal{U}(2)$ &5.40 &4.40 &4.60 &5.20 &4.80 &5.50 &5.50  \cr
$\mathcal{U}(3)$ &5.10 &5.10 &4.40 &5.50 &5.60 &4.80 &5.40  \cr
$\mathcal{U}(4)$ &5.10 &6.20 &7.40 &6.40 &6.10 &7.00 &4.10  \cr
$\mathcal{U}(5)$ &4.80 &6.10 &5.10 &5.70 &4.70 &5.80 &4.60  \cr
$\mathcal{U}(6)$ &3.20 &3.90 &4.80 &6.20 &6.10 &6.70 &5.40  \cr
QC &4.50 &3.90 &4.90 &5.00 &4.90 &5.90 &6.20  \cr
XW &4.40 &4.00 &5.50 &6.00 &3.70 &4.90 &5.50  \cr
\hline
\end{tabular}}
\end{center}
\end{table}

\begin{table}[h]
\caption{Empirical sizes under Setting 1 for multivariate $t_7(\0,\bSig)$ and $n = 100$ (in percentage).}
\renewcommand{\arraystretch}{0.9}\doublerulesep 0.01pt\tabcolsep 0.18in
\begin{center}
\label{size4}
\begin{tabular}{cp{0.5cm}<{\centering}p{0.5cm}<{\centering}p{0.5cm}<{\centering}p{0.5cm}<{\centering}p{0.5cm}<{\centering}p{0.5cm}<{\centering}p{0.5cm}<{\centering}}
\hline
$p$  & 50 & 100 & 200 & 400 & 600 & 800 &1000 \cr
\hline
$\mbox{adpUmin}$ &6.50 &6.10 &6.10 &6.00 &8.30 &5.80 &7.20  \cr
$\mbox{adpUf}$ &8.00 &7.00 &5.70 &5.60 &7.70 &5.80 &6.50  \cr
$\mathcal{U}(1)$ &3.80 &4.60 &4.30 &5.40 &5.70 &5.00 &4.40  \cr
$\mathcal{U}(2)$ &6.00 &6.30 &5.50 &4.40 &5.90 &5.90 &6.60  \cr
$\mathcal{U}(3)$ &6.00 &5.00 &4.70 &6.10 &5.30 &4.30 &5.00  \cr
$\mathcal{U}(4)$ &5.20 &6.00 &5.10 &5.30 &5.90 &5.50 &5.40  \cr
$\mathcal{U}(5)$ &5.40 &5.90 &5.70 &6.00 &5.80 &5.10 &6.40  \cr
$\mathcal{U}(6)$ &4.30 &5.20 &4.90 &4.50 &5.30 &5.90 &5.60  \cr
QC &17.4 &18.5 &20.3 &19.6 &19.2 &21.5 &19.6  \cr
XW &1.20 &1.60 &0.80 &0.80 &0.90 &0.80 &0.70 \cr
\hline
\end{tabular}
\end{center}
\end{table}

\begin{figure}[htbp]
\centering
\subfigure{
\includegraphics[width=2.4in]{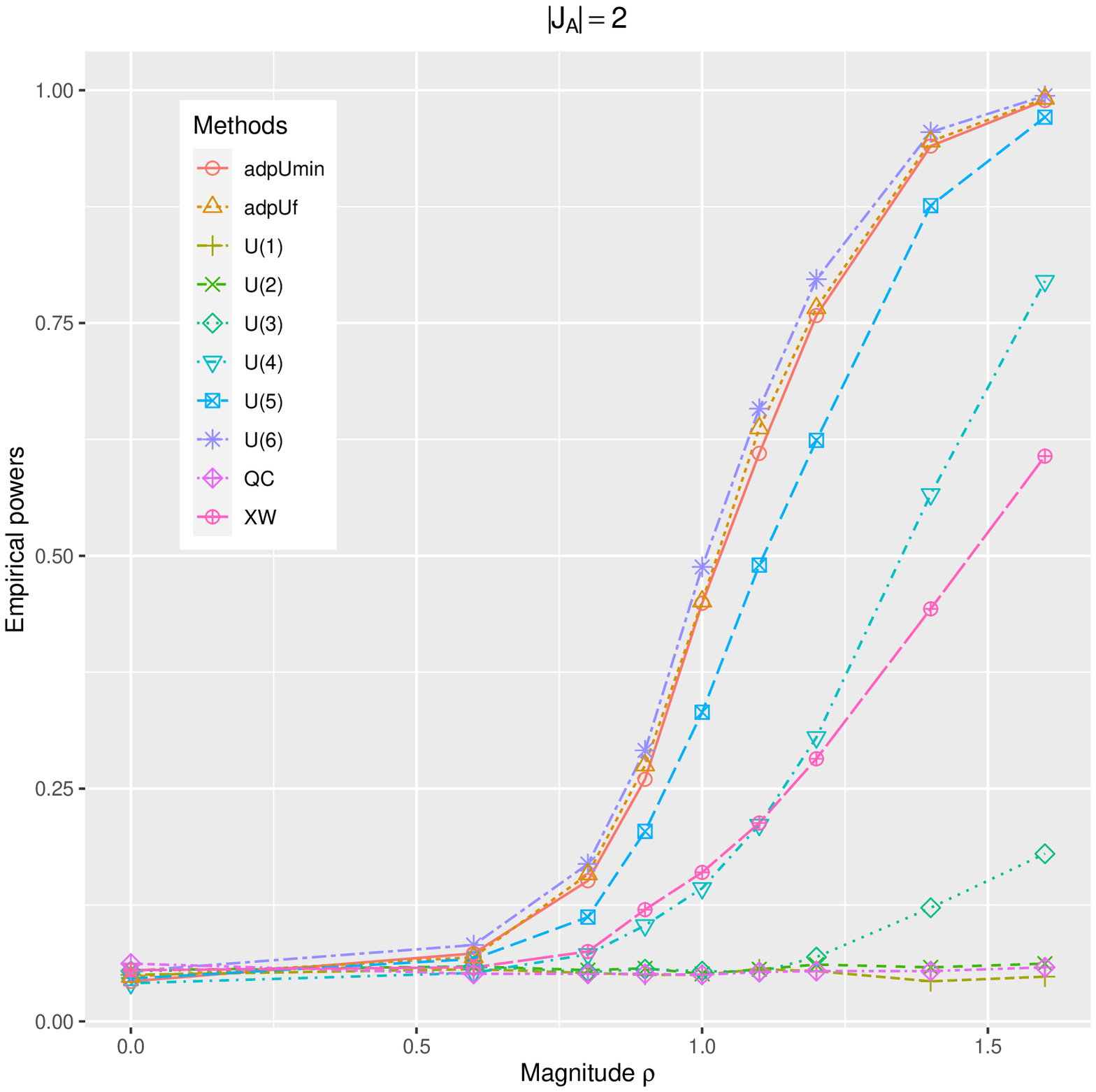}
          }
\subfigure{
\includegraphics[width=2.4in]{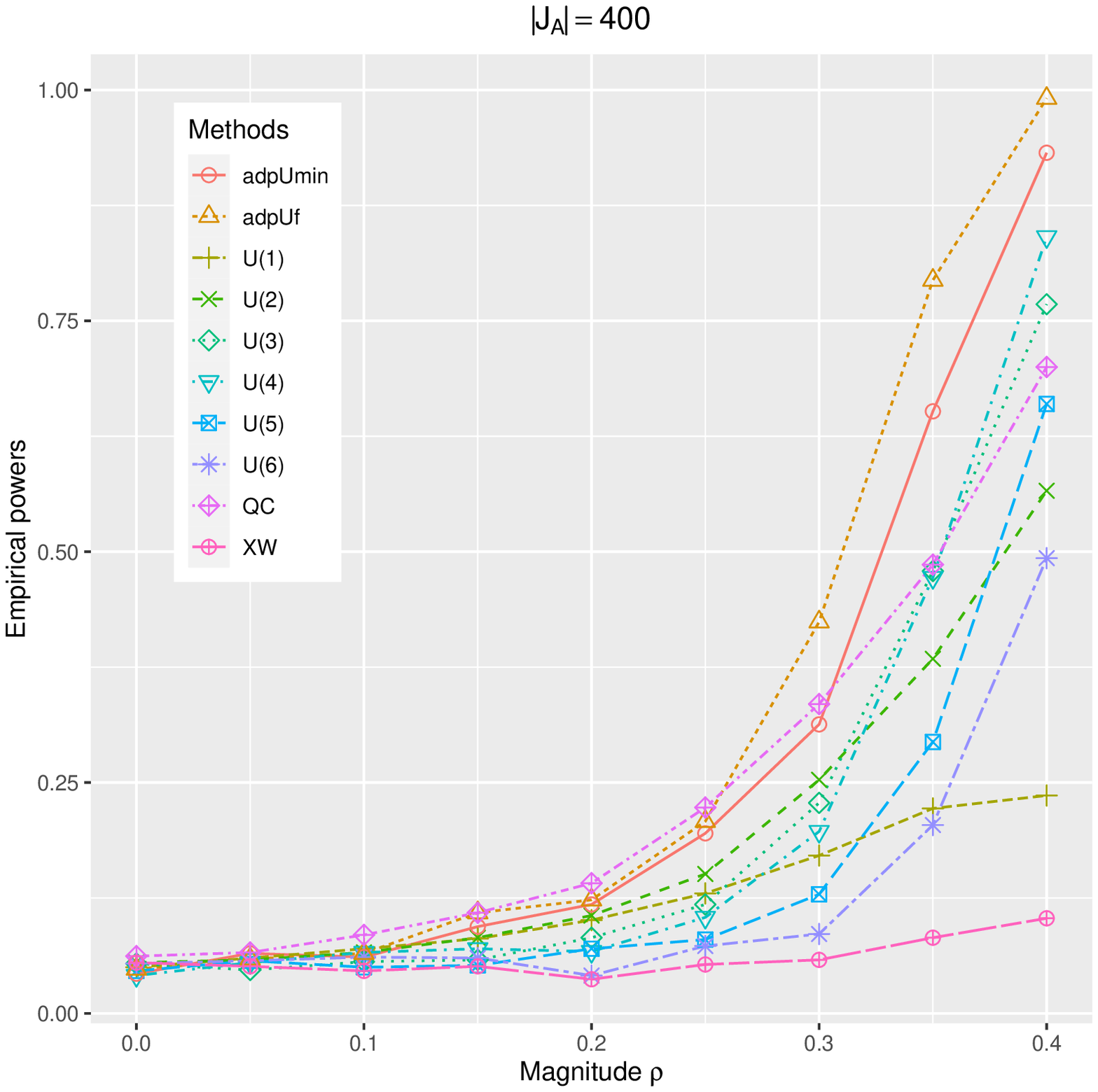}
}
\vspace{-4mm}
\subfigure{
\includegraphics[width=2.4in]{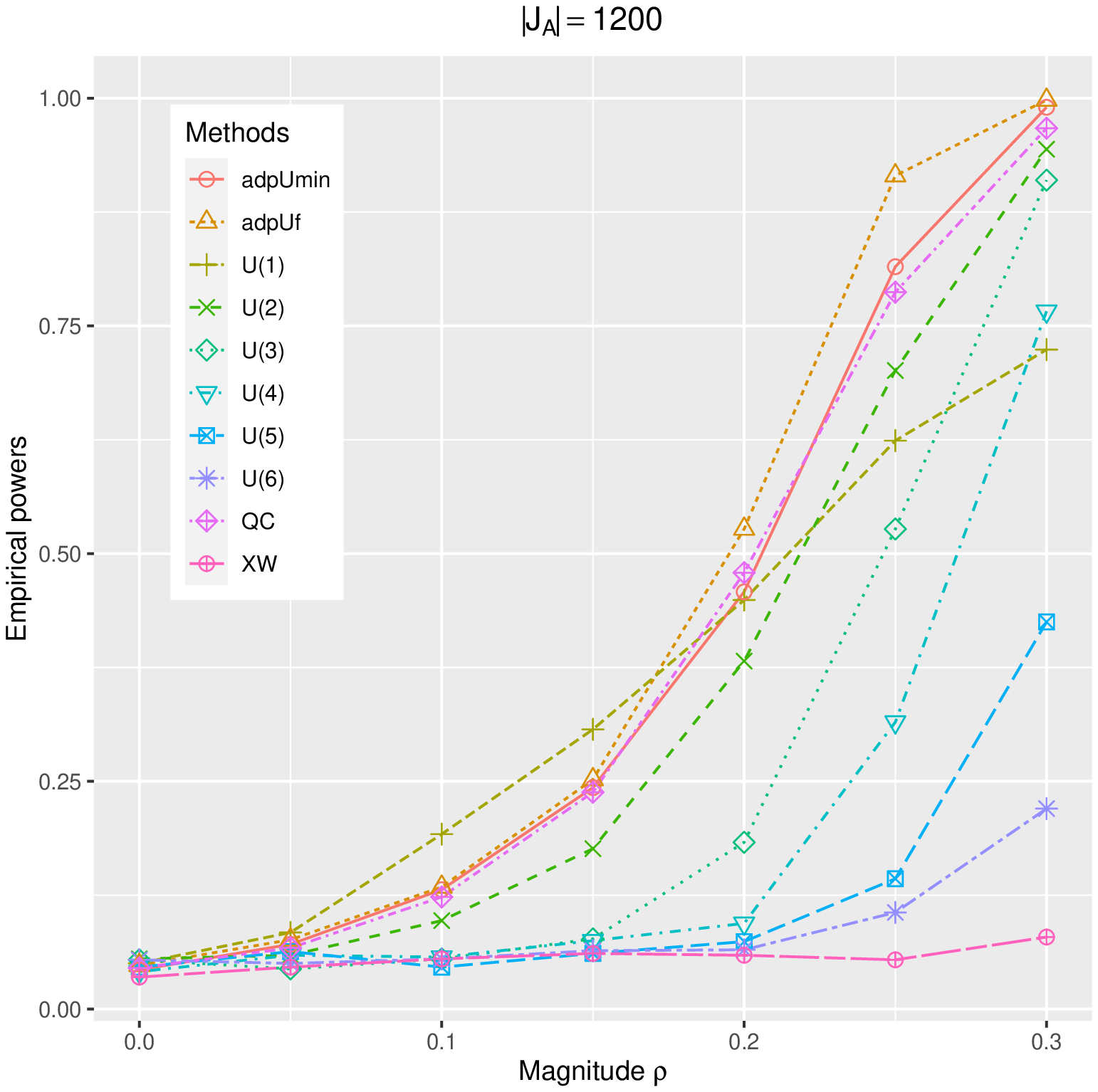}
}
\subfigure{
\includegraphics[width=2.4in]{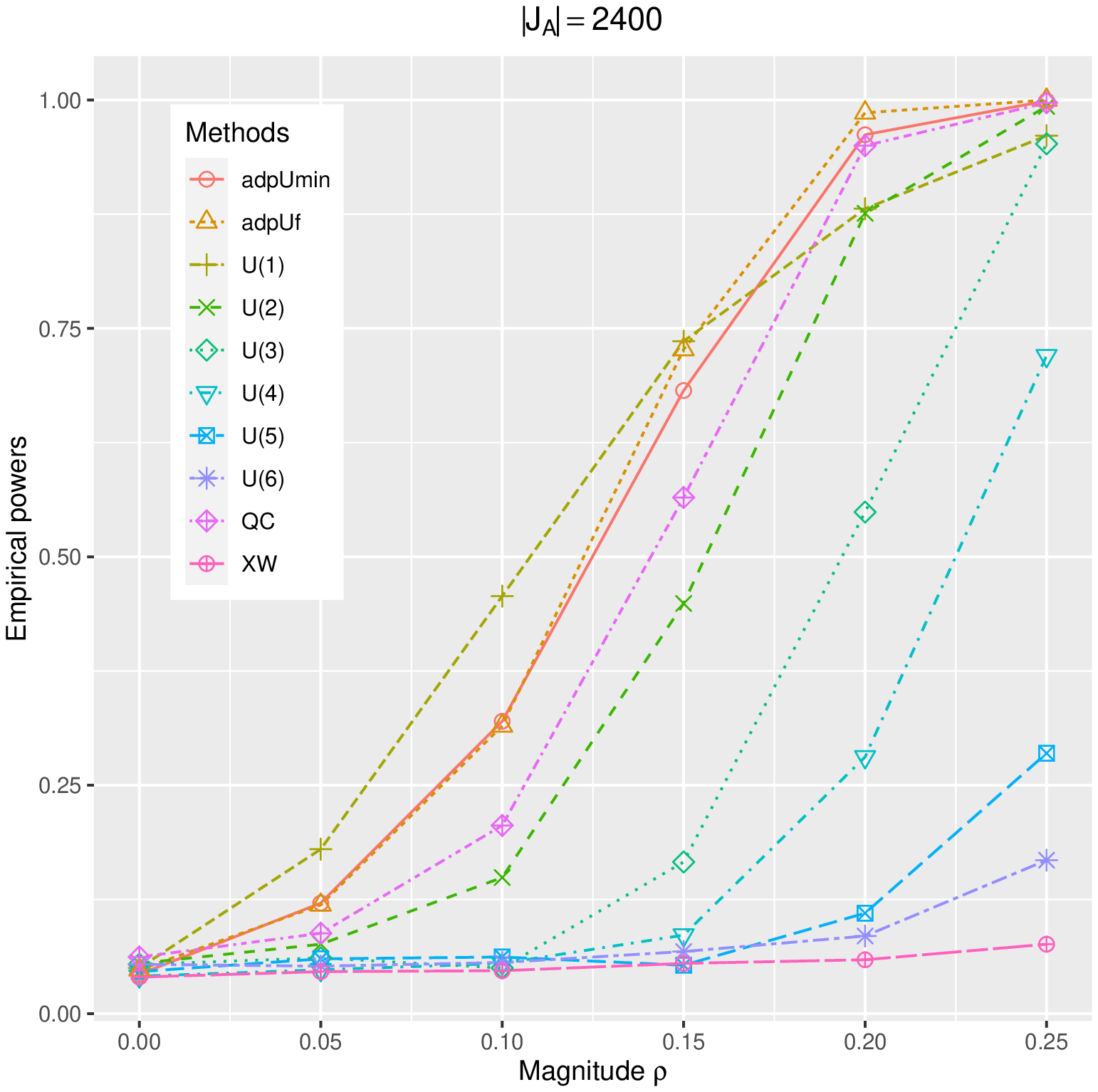}
}
\centering
\caption{Empirical power comparison under Setting 1 for multivariate normal distribution: $n=100$, $p=1000$.}
\label{fig1}
\end{figure}

\begin{table}[htbp]
\caption{Averaged empirical bias and standard deviation in parentheses of three bandwidth estimators with normal innovations: our proposed adaptive bandwidth estimator with $\delta=0.5$ and $\theta=0.06$, the estimators proposed in \cite{bickel2008regularized} (BLa) and \cite{qiu2012test} (QC).}
\label{band-estimation-normal}
\doublerulesep 0.5pt
\begin{center}
\begin{tabular*}{\textwidth}{p{.4cm}<{\centering}p{.45cm}<{\centering}p{1.3cm}<{\centering}cccc}
\hline
 & & & \multicolumn{4}{c}{Bandwidth} \cr
\cmidrule{4-7}
$n$ & $p$ & Methods   & 2 & 5 & 10 & 15 \cr
\hline
100 & 50 & Adaptive  &0.04(0.243) &0(0) &0(0) &-0.03(0.171)  \cr
&  & BLa  &0.15(0.411) &-0.37(0.691) &-0.89(1.144) &-0.96(1.809)  \cr
&  & QC  &0(0) &0(0) &0(0) &-0.02(0.141)  \cr
& 200 & Adaptive  &0(0) &0(0) &0(0) &0(0)   \cr
&  & BLa  &0.38(0.663) &0.27(1.062) &0.14(1.128) &-0.44(1.641)\cr
&  & QC  &0(0)  &0(0) &0(0) &0(0)  \cr
& 400 & Adaptive  &0(0) &0(0) &0(0) &0.01(0.100)   \cr
&  & BLa &0.56(1.258) &0.84(1.631) &0.50(1.554) &0.12(1.653)\cr
&  & QC  &0(0) &0(0) &0(0) &0(0)  \cr
& 600 & Adaptive  &0(0) &0(0) &0(0) &0(0)   \cr
&  & BLa  &0.91(1.518) &0.74(1.574) &0.41(1.615)  &0.59(2.216)\cr
&  & QC  &0(0) &0(0) &0(0) &0(0)   \cr
& 1000 & Adaptive  &0(0) &0(0) &0(0)  &0(0)  \cr
&  & BLa &1.61(2.260) &1.44(2.328) &1.22(2.389) &0.69(2.862)  \cr
&  & QC  &0(0) &0(0) &0(0) &0(0)  \cr
200 & 50 & Adaptive &0(0) &0.01(0.100) &0.04(0.243) & 0.04(0.315) \cr
&  & BLa  &0.11(0.345) &0.09(0.637) &0.19(0.929) &0.22(1.630)\cr
&  & QC &0.01(0.100) &0.01(0.100) &0(0) &0(0)   \cr
& 200 & Adaptive &0(0) &0(0)  &0(0) &0(0) \cr
&  & BLa &0.29(0.537) &0.34(0.879) &0.14(0.899) &0.05(1.290)  \cr
&  & QC &0(0) &0(0) &0(0) &0(0)   \cr
& 400 & Adaptive &0(0)  &0(0)  &0(0) &0(0)   \cr
&  & BLa &0.71(1.028) &0.70(1.087) &0.44(1.122) &0.43(1.513)  \cr
&  & QC &0(0) &0(0) &0(0) &0(0)  \cr
& 600 & Adaptive &0(0)  &0(0)  &0(0)  &0(0)  \cr
&  & BLa  &0.88(1.17) &1.02(1.463) &0.75(1.344) &1.14(1.809)  \cr
&  & QC &0(0) &0(0) &0(0) &0(0)   \cr
& 1000 & Adaptive &0(0) &0(0) &0(0) &0(0)  \cr
&  & BLa  &1.23(1.399) &1.50(1.957) &1.15(1.822) &1.41(2.396)   \cr
&  & QC &0(0) &0(0) &0(0) &0(0)   \cr
\hline
\end{tabular*}
\end{center}
\end{table}

\section{Data Analysis}
\label{Section-4}
In this section, we applied our proposed procedures to a prostate cancer data set from a protein mass spectroscopy study \citep{adam2002serum}, which analyzed the constituents of the proteins in the blood for two groups of people -- the healthy group and the cancer group.
The data set has also been  studied in \cite{levina2008sparse} and  \cite{qiu2012test}.
For each blood serum sample $i$,  the data consist of the intensity $X_{ij}$ for a large number of time-of-flight
values $t_j$, which is related to the mass over charge ratio of the
constituent proteins.
We analyzed the standardized data set, which consists of  157 healthy and 167 cancer patients,  with a 218-dimensional intensity vector for each individual. 

We focus on testing a string of null hypotheses $H_{k,0}: \bSig=\boldsymbol{B}_k(\bSig)$, $k=0,1,\dots,216$ and estimating the bandwidth of the covariance matrices of the healthy and cancer groups. In particular, we choose $\delta=0.5$ and $\theta=0.005$ in analyzing the real data with our adaptive estimator, which is consistent with the choice of \cite{qiu2012test}.
We exhibit some representative $p$-values in Table \ref{realdata-1} and bandwidth estimates in Table \ref{realdata-2}.


All $p$-values of  ``adpUmin" and ``adpUf" tests are very close to zero borrowing strength from $\mathcal{U}(5)$ and $\mathcal{U}(6)$, and the estimated values of our proposed adaptive estimator are 203 and 216 for the healthy group and cancer group. In practice, a covariance matrix with large bandwidth may not be valuable because it will not significantly reduce the number of parameters. Thus, these small $p$-values and bandwidth estimates suggest that the covariances of both the healthy group and cancer group may not be banded. In the meanwhile, the heatmaps in Figure \ref{realdata-fig2} display that most of the sample correlations in the whole matrices of the healthy group and the cancer group are non-negligible, leading to non-banded structures. It also supports our conclusion. 

\begin{sidewaystable}[htbp]
\caption{\rm The p-values (\%) of various tests applied to the prostate cancer data set.}
\renewcommand{\arraystretch}{0.8}\doublerulesep 0.03pt\tabcolsep 0.06in
\begin{center}
\begin{threeparttable}
\begin{tabular}{cc cccccccccc}
\hline
&& \multicolumn{9}{c}{Bandwidith}  \cr
\cmidrule{3-11}
&  Test & 5 & 86 & 106 & 109 & 116 &125 & 150 & 200 & 216   \cr
\cmidrule{2-11}
Health  & \mbox{adpUmin} &0 &0 &0 &0 &0 &0 &0 &0 &0   \cr
Group  &\mbox{adpUf} &0 &0 &0 &0 &0 &0 &0 &0 &0    \cr
& $\mathcal{U}(1)$ &0 &6.03 &99.54 &75.87 &32.32 &8.39& 0.19 &1.21e-03 &1.08e-03    \cr
& $\mathcal{U}(2)$ &0 &$<$1.0e-13 &$<$1.0e-13 &$<$1.0e-13 &1.4e-07 &4.47e-6& 1.7e-12 &$<$1.0e-13 &$<$1.0e-13    \cr
& $\mathcal{U}(3)$ &0 &$<$1.0e-13 &16.71 &16.87 &2.4e-09 &$<$1.0e-13 &$<$1.0e-13 &$<$1.0e-13 &$<$1.0e-13   \cr
& $\mathcal{U}(4)$ &0 &0 &0 &$<$1.0e-13 &$<$1.0e-13 &$<$1.0e-13 &$<$1.0e-13 &0 &0    \cr
& $\mathcal{U}(5)$ &0 &0 &$<$1.0e-13 &2.4e-03 &$<$1.0e-13 & $<$1.0e-13 &0 &0 &0     \cr
& $\mathcal{U}(6)$ &0 &0 &0 &0 &0 &0 &0 &0 &0    \cr
& QC &0 &2.06e-7 &2.06e-7 &0 &5.64 &10.36 &12.92 &42.69 &49.95  \cr
& XW &0 &0 &3.4e-07 &4.9e-06 &1.6e-03 &2.2e-2 &2.2e-2 &3.1e-2 &8.5e-2   \cr
\hline
&& \multicolumn{9}{c}{Bandwidith}  \cr
\cmidrule{3-11}
&  Test & 5 &  52 & 61 & 67 & 91  & 150 & 191 & 200 & 216   \cr
\cmidrule{2-11}
Cancer &\mbox{adpUmin} &0 &0 &0 &0 &0 &0 &0 &0 &0   \cr
Group &\mbox{adpUf} &0 &0 &0 &0 &0 &0 &0 &0 &0    \cr
& $\mathcal{U}(1)$ &0 &6.19 &45.20 &96.22 &3.3e-01 &2.8e-10 &4.1e-10 &4.1e-09 &2.4e-07    \cr
& $\mathcal{U}(2)$ &0 &$<$1.0e-13 &$<$1.0e-13 &$<$1.0e-13 &$<$1.0e-13 &$<$1.0e-13 &$<$1.0e-13 &$<$1.0e-13 &$<$1.0e-13    \cr
& $\mathcal{U}(3)$ &0 &$<$1.0e-13 &19.85 &$<$1.0e-13 &0 &0 &0 &0 &0    \cr
& $\mathcal{U}(4)$ &0 &0 &0 &0 &0 &0 &0 &0 &0    \cr
& $\mathcal{U}(5)$ &0 &0 &0 &0 &0 &0 &0 &0 &0    \cr
& $\mathcal{U}(6)$ &0 &0 &0 &0 &0 &0 &0 &0 &0    \cr
& QC  &0 &0 &0 &0 &0 &1.3e-8 &5.66 &22.52 &49.79 \cr
& XW &0 &0 &6.0e-12 &2.7e-11 &2.7e-11 &8.0e-11 &8.0e-11 &8.0e-11 &7.8e-07   \cr
\hline
\end{tabular}
\begin{tablenotes}
        \footnotesize
        \item $\mathcal{U}(a)$, the proposed tests based on U-statistics with different values of $a$;
        \mbox{adpUmin}, the adaptive test based on the minimax method;
        \mbox{adpUf}, the adaptive test based on the Fisher combination;
         QC, test of \cite{qiu2012test};
         XW, test of \cite{xiao2013asymptotic}.
 \end{tablenotes}
\end{threeparttable}
\end{center}
\label{realdata-1}
\end{sidewaystable}


\begin{figure}[H]
\subfigure{
\includegraphics[width=2in]{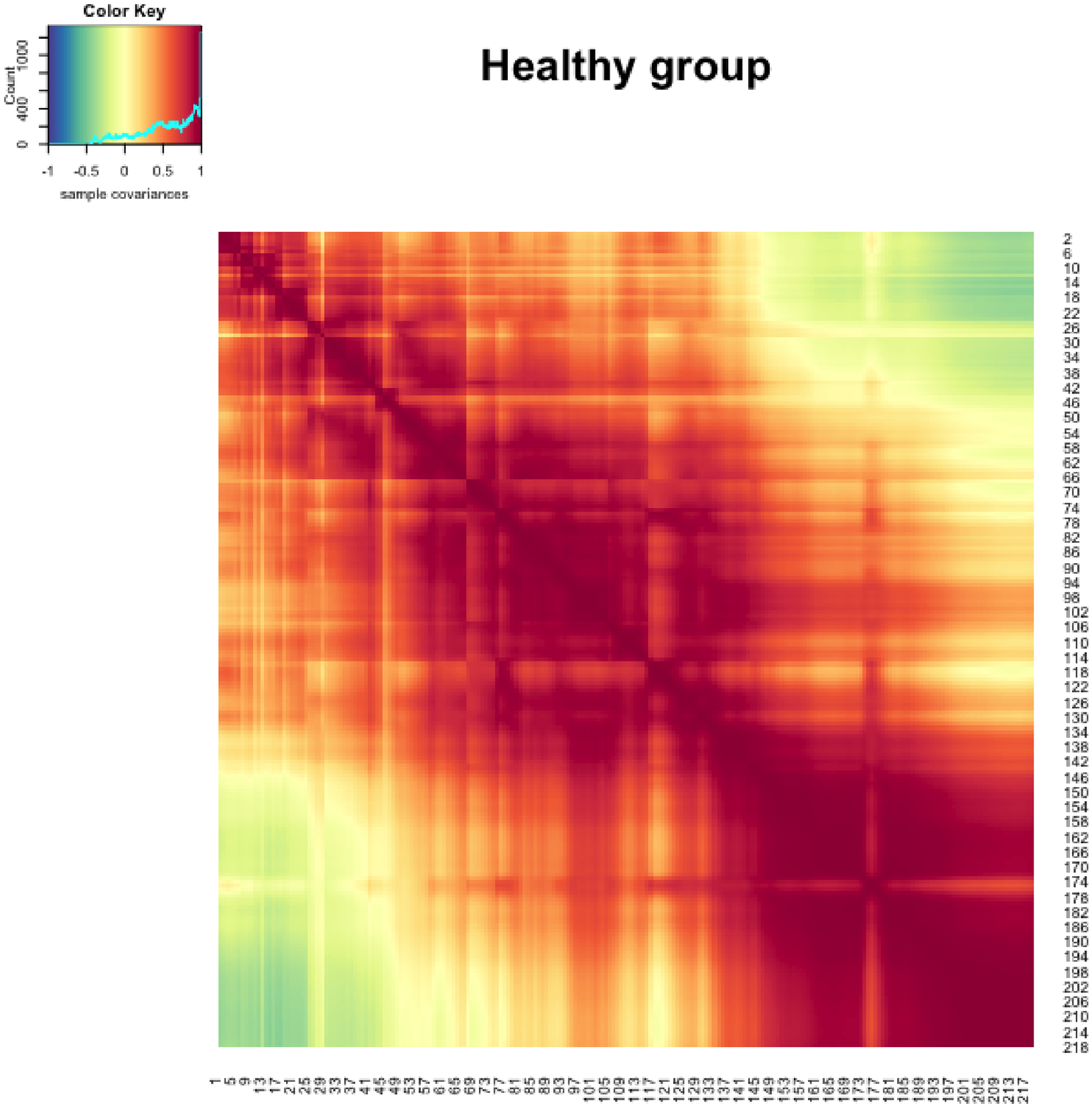}
}
\subfigure{
\includegraphics[width=2in]{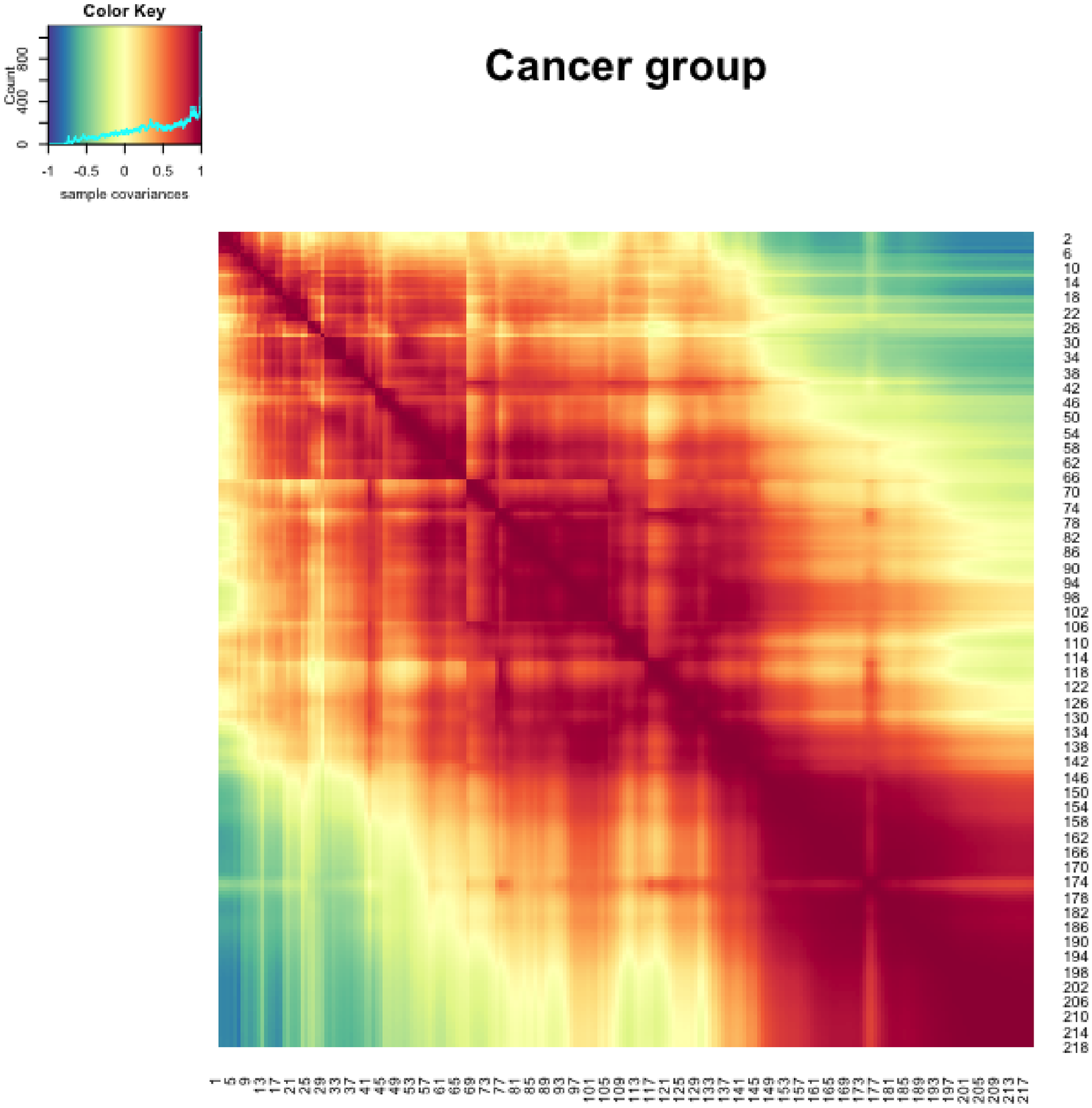}
}
\centering
\caption{Heatmaps for the sample covariance matrices of the healthy and the cancer group. Blue represents negative correlation, red represents positive correlation, and the color deepens as the correlation increases.}
\label{realdata-fig2}
\end{figure}

\begin{table}[htb]
\caption{\rm The estimated bandwidths of various procedures applied to the prostate cancer data set.}
\label{realdata-2}
\begin{center}
\small
\begin{tabular}{ ccccccccc }
\hline
 Method &  $\mathcal{U}(1) $ & $\mathcal{U}(2)$ & $\mathcal{U}(3)$ & $\mathcal{U}(4)$  & $\mathcal{U}(5)$ & $\mathcal{U}(6)$ & Adaptive & QC\cr
\cmidrule{2-9}
Healthy Group&132 &120 &193 &123 &203 &128 &203 &121 \cr
Cancer Group&120  &74  &171 &173 &175 &216 & 216 &212 \cr
\hline
\end{tabular}
\end{center}
\end{table}

 A similar conclusion is obtained with the XW test as they also provided very small $p$-values for all hypotheses.
However, the QC test gave different conclusions, where the smallest $k$ such that $H_{k,0}$ is not rejected is 116  for the healthy group, while is 191 for the cancer group.
Note that the statistic values of $\mathcal{U}(2)$ test are as same as that of QC test. We found that
the performance of the QC test is different from $\mathcal{U}(2)$ test with large $k$. One possible reason is that the variance estimation of the statistic in QC test is based on the assumption that $\tr(\bSig^4)/\tr^2(\bSig^2)=O(p^{-1})$, which may not be satisfied by the real data, while our method does not rely on such an assumption to estimate the variance of $\mathcal{U}(2)$.
To explain the difference between $\mathcal{U}(2)$ test and QC test, we considered a special case with testing $H_{k,0}: \bSig = B_k(\bSig)$, where $\bSig=\bGamma\bGamma^{\T}$ and $\bGamma=(\gamma_{j_1 j_2})_{p\times p}$ with $\gamma_{j_1 j_1}=1$, $\gamma_{j_1 j_2}\sim \mbox{Unif(0,5)}$ for $0<j_1-j_2\leq k$ with $k = 5$ and $200$. We generated 1000 datasets with the sample size $n=157$ and the dimension $p=218$ from $N(\mathbf{0},\bSig)$ under $H_{k,0}$ .
Table \ref{realdata-sim} shows that the ASD and AEASD of the QC test are too far away from the MCSD, the type I error of the QC test is therefore very small compared to the nominal level $5\%$ when $k=200$. 
In this scenario, $\tr(\bSig^4)/\tr^2(\bSig^2)=0.944$ and $1000^{-1}\sum\nolimits_{b=1}^{1000}\tr(\bS_{n,b}^4)/\tr^2(\bS_{n,b}^2)=0.936$, where $\bS_{n,b}=(n-1)^{-1}\sum\nolimits_{i=1}^{n}(\bx_i^{(b)}-\bar{\bx}^{(b)})(\bx_i^{(b)}-\bar{\bx}^{(b)})^{\T}$ with $\bar{\bx}^{(b)}=n^{-1}\sum\nolimits_{i=1}^{n}\bx_i^{(b)}$, 
and $\{\bx_1^{(b)},\dots,\bx_n^{(b)}\}$ is the $b$-th sampling from $N(\mathbf{0},\bSig)$.
It indicates that the QC test may give an overestimation of the variance of the statistic when $\tr(\bSig^4)/\tr^2(\bSig^2)$ is large and thus violates their assumption. QC test still works well when $\tr(\bSig^4)/\tr^2(\bSig^2)$ is small, e.g. $\tr(\bSig^4)/\tr^2(\bSig^2)=0.024$ and $1000^{-1}\sum\nolimits_{b=1}^n \tr(\bS_{n,b}^4)$ $/\tr^2(\bS_{n,b}^2)=0.041$ with $k=5$. In the real data analysis, for the health and cancer groups, $\tr(\bS_{n}^4)/\tr^2(\bS_{n}^2)=0.907$ and 0.820, respectively, indicating a possible overestimation of the variance of the QC test statistic.

\begin{table}[h]
\caption{Results based on 1000 multivariate normal samples under $H_{k,0}: \bSig = B_k(\bSig)$.}
\label{realdata-sim}
\begin{center}
\small
\begin{tabular}{ccccc}
\hline
& \multicolumn{4}{c}{$\mbox{Case 1}: k=5$} \cr
 &  ASD & AEASD & MCSD & Type I Error  \cr
\cmidrule{2-5}
$\mathcal{U}(2)$& 20498.76 & 20242.96 &21317.02&  0.069 \cr
QC&21059.26 & 21079.46 &21317.02 & 0.054 \cr
\hline
& \multicolumn{4}{c}{$\mbox{Case 2}: k=200$} \cr
 &  ASD & AEASD & MCSD & Type I Error  \cr
\cmidrule{2-5}
$\mathcal{U}(2)$&96270.76& 92689.35 & 94453.38 & 0.046 \cr
QC & 152093323 & 151994013 & 94453.38 & 0 \cr
\hline
\end{tabular}
\begin{tablenotes}
        \footnotesize 
        \item 
        ASD, asymptotic standard deviation of statistic;  
        AEASD, average of estimations \\of the asymptotic standard deviation of the statistics based on 1000 replications;
        \\MCSD, sample standard deviation of the statistics based on 1000 replications.
\end{tablenotes}
\end{center}
\end{table}

\section{Discussion}
\label{Section-5}
In this paper, we propose adaptive tests based on a series of U-statistics for testing the bandedness of the high-dimensional covariance matrix. 
We investigate the asymptotic joint distribution of the U-statistics under the null hypothesis and specific local alternative hypotheses.
Further, we take advantage of the asymptotic independence of multiple U-statistics to construct two proposed adaptive tests by combining the $p$-values of U-statistics. 
The simulation studies show that the proposed tests are powerful across a wide range of alternatives, whereas the existing tests are only powerful for either dense alternatives or sparse alternatives.
We also propose a new consistent bandwidth estimator motivated by the by-product of the U-statistics.

The bandwidth $k$ is usually unknown in practice. Instead of testing a general bandedness structure of covariance matrix with a given $k$, it is of great interest to regard the bandwidth $k$ as a tuning parameter and examine the asymptotic properties of a series U-statistics based on $\sum\nolimits_{\hat{k}<|j_1-j_2|< p}\sigma_{j_1j_2}^a$, where $\hat{k}$ is the estimation of the true bandwidth $k$.  As shown in \cite{zhong2017tests}, the plug-in estimator $\hat{k}$ may incur some leading order effects. We will examine this topic in the future.

\vskip 14pt
\noindent {\large\bf Supplementary Materials}

 The Supplementary Materials contain detailed proofs of the theoretical results and more simulation results.
\par

\section*{Acknowledgement}
We would like to thank the editor, the associate editor, and two anonymous reviewers for their constructive comments, which led to substantial improvements. We are also grateful to Dr. Yumou Qiu for sharing the prostate cancer data set used in \cite{qiu2012test}. Shurong Zheng and Xiaoyi Wang were supported by NSFC grant 12071066 and Gongjun Xu was supported by NSF grant  SES 1846747.

\par

\markboth{\hfill{\footnotesize\rm FIRSTNAME1 LASTNAME1 AND FIRSTNAME2 LASTNAME2} \hfill}
{\hfill {\footnotesize\rm FILL IN A SHORT RUNNING TITLE} \hfill}

\bibhang=1.7pc
\bibsep=2pt
\fontsize{9}{14pt plus.8pt minus .6pt}\selectfont
\renewcommand\bibname{\large \bf References}
\expandafter\ifx\csname
natexlab\endcsname\relax\def\natexlab#1{#1}\fi
\expandafter\ifx\csname url\endcsname\relax
  \def\url#1{\texttt{#1}}\fi
\expandafter\ifx\csname urlprefix\endcsname\relax\def\urlprefix{URL}\fi

\lhead[\footnotesize\thepage\fancyplain{}\leftmark]{}\rhead[]{\fancyplain{}\rightmark\footnotesize{} }

\bibliographystyle{apalike}
\bibliography{xiaoyi}

\vskip .65cm
\noindent
Center for Statistics and Data Science, Beijing Normal University, Zhuhai 519087, China.
\vskip 2pt
\noindent
E-mail: wangxy059@nenu.edu.cn
\vskip 2pt

\noindent
Department of Statistics, University of Michigan, Ann Arbor, MI 48109, USA.
\vskip 2pt
\noindent
E-mail: gongjun@umich.edu
\vskip 2pt

\noindent
School of Mathematics \& Statistics and KLAS, Northeast Normal University, Changchun 130024, China
\vskip 2pt
\noindent
E-mail: zhengsr@nenu.edu.cn

\end{document}